\documentclass[11pt]{article}
\usepackage{arydshln}
\usepackage{color}
\usepackage[body={16cm,23cm}]{geometry}
\usepackage{cite}
\usepackage[nottoc]{tocbibind}
\usepackage{amssymb,amsmath,amsthm,amsopn}
\usepackage{amsfonts}
\usepackage{mathabx}
\usepackage[mathscr]{eucal}
\usepackage{epic}
\usepackage{eepic}
\usepackage{enumerate}
\usepackage{graphicx}
\usepackage[small,bf%,nooneline
]{caption}
\makeatletter
\@addtoreset{equation}{section}
\makeatother
\renewcommand{\theequation}{\thesection.\arabic{equation}}

\newcounter{app}
\newcounter{sapp}[app]
\def\theapp{\Alph{app}}
\newcommand{\app}[1]{
\refstepcounter{app}{\vspace{7mm}
\noindent\Large\bf Appendix
\theapp.
 \ #1 \par \vspace{5mm}}
\setcounter{equation}{0}
\def\theequation{\Alph{app}.\arabic{equation}}}

\newcommand{\ds}{\displaystyle}
\def\EXP{\textrm{{\large e}}}
\newcommand{\xop}{\mathbf{x}}
\newcommand{\pop}{\mathbf{p}}
\newcommand{\R}{\mathbf{R}}

\newcommand{\ii}{\mathsf{i}}

\newcommand{\Rcal}{{\mathcal R}}
\newcommand{\be}{\begin{equation}}
\newcommand{\beq}{\begin{equation}}
\newcommand{\ee}{\end{equation}}
\newcommand{\eeq}{\end{equation}}
\newcommand{\uf}{{\mathbf u}}
\newcommand{\vf}{{\mathbf v}}
\newcommand{\gf}{{\mathbf g}}

\newcommand{\bb}{\mathsf{b}}
\newcommand{\Rbar}{\overline{\Rcal}}

\newcommand{\dbar}{\overline{\delta}}
\newcommand{\Dbar}{\overline{\Delta}}
\newcommand{\Sbar}{\overline{S}}
\newcommand{\ebar}{\overline{\epsilon}}
\newcommand{\U}{\mathcal{U}}
\newcommand{\mtimes}{\,\dot\otimes\, \,}
\newcommand{\vark}{s}

\newcommand\X{{\mathcal X}}
\renewcommand{\author}[1]{\large\rm #1\\ \bigskip}
\newcommand{\address}[1]{{\normalsize\it #1\\}\bigskip}
\renewcommand{\title}[1]{\bigskip\bigskip\Large\bf #1\bigskip\bigskip\\}

\def\A{{\mathcal A}}
\def\R{\mathcal{R}}
\def\P{{\mathcal P}}

\def\Rb{{\mathbf R}}

\newcommand{\nocontentsline}[3]{}
\newcommand{\tocless}[2]{\bgroup\let\addcontentsline=\nocontentsline#1{#2}\egroup}

\begin{document}

\vglue 2cm

\begin{center}

\title{Yang-Baxter Maps, Discrete Integrable Equations\\ and Quantum Groups}

\vspace{.5cm}

\author{Vladimir V.~Bazhanov$^{1}$ and Sergey M.~Sergeev$^{1,2}$}

\vspace{.5cm}

\address{$^1$Department of Theoretical Physics,
         Research School of Physics and Engineering,\\
    Australian National University, Canberra, ACT 2601, Australia.\\\ \\
$^2$Faculty of Education Science Technology \& Mathematics,\\
University of Canberra, Bruce ACT 2601, Australia.}

\abstract{For every quantized Lie algebra there exists a
  map from the tensor square of the algebra to itself, which by construction
  satisfies the set-theoretic Yang-Baxter equation. This map allows
  one to define an integrable discrete 
  quantum evolution system on quadrilateral lattices, where local
  degrees of freedom (dynamical variables) take values in a tensor
  power of the quantized Lie algebra. The corresponding equations of
  motion admit the zero curvature representation. The commuting Integrals
  of Motion are defined in the standard way via the Quantum 
  Inverse Problem Method,  
  utilizing Baxter's famous commuting transfer matrix
  approach.
All elements of the above construction have a meaningful
quasi-classical limit.  As a result one obtains an integrable discrete
Hamiltonian evolution system, where the local equation of motion are
determined by a classical Yang-Baxter map and the action functional is
determined by the quasi-classical asymptotics of the universal
$\Rb$-matrix of the underlying quantum algebra.
In this paper we present detailed considerations of the above scheme
on the example of the algebra $U_q(sl(2))$ leading to discrete
Liouville equations, however the approach is rather general and can be
applied to any quantized Lie algebra.}

\end{center}

%\maketitle

%-----------------------------------------------------

\vspace{1cm}
\newpage
\vspace {1cm}
\tableofcontents

%------------------------------------------------------
\vspace{1cm}
\newpage
%\vglue 1.cm
\section{Introduction}

The ``Yang-Baxter maps'' \cite{Veselov:2003} are invertible maps of
a Cartesian product of two identical sets $\X$,  
\beq
\R:\qquad \X\times\X\mapsto\X\times\X\label{R-def}
\eeq
satisfying the ``functional'' or ``set-theoretic'' 
 Yang-Baxter equation \cite{Drinfeld:1992}, 
\beq
\R_{12}\circ\R_{13}\circ\R_{23}=\R_{23}\circ\R_{13}\circ\R_{12}\,.\label{SYBE}
\eeq
This equation states an equality of two different composition of 
the three maps $\R_{12}$, $\R_{13}$ and $\R_{23}$, acting on different  
factors in a product of three sets
$\X\times\X\times\X$ 
(for instance, $\R_{12}$ coincides with the map \eqref{R-def}
in the first and second factors and acts as an identity map in the third one;
the maps $\R_{13}$ and $\R_{23}$ are defined similarly).

The interest to the Yang-Baxter maps is motivated mostly by their
connection to discrete integrable evolution equations.  Important examples of
the Yang-Baxter maps as well as some classification results 
were obtained in \cite{Adler:1993,Hietarinta:1997,Etingof:1999,Quispel:1999,
Joshi:2010,
Kajiwara:2002,Odesskii:2003,Goncharenko:2004,Adler:2004}. Mention also a
related {\em consistency around a cube} condition \cite{AdlerBobenkoSuris}, 
which is in many cases can be associated with some set-theoretic Yang-Baxter
equation. One common limitation of the existing methods
is that Hamiltonian structures of the arising maps generally remain
unclear. As a result there are no regular procedures for 
quantization of the Yang-Baxter maps or their applications to 
Hamiltonian evolution systems.
    
In this paper we address these problems and present a new approach to
the Yang-Baxter maps, which is based on the theory of quantum groups
\cite{Dri87,Jim86} 
and naturally connected to Baxter's commuting transfer matrices
\cite{Bax72} and the Quantum Inverse Problem Method
\cite{Faddeev:1979gh}.
The key role in our approach will be played
by the universal $\Rb$-matrix \cite{Dri87}. This notion is associated
with the so-called, {\em quasitriangular Hopf algebras}, which include
a very important class of the $q$-deformed (affine) Lie algebras and
super-algebras (more precisely, their universal enveloping algebras).
Let $\A$ be a quasitriangular Hopf algebra, then there exists an
invertible element $\Rb\in\A\otimes\A$, belonging to the tensor
product of two algebras $\A$, called the universal $\Rb$-matrix, which
by construction satisfies the quantum Yang-Baxter equation
\beq
\Rb_{12}\ \Rb_{13}\ \Rb_{23}=\Rb_{23}\ \Rb_{13}\ \Rb_{12}\,,\qquad
\Rb\in \A\otimes\A\,.\label{YBE}
\eeq
Similarly to Eq.\eqref{SYBE} the indices here indicate how the
universal $\Rb$-matrix is embedded into the triple product
$\A\otimes\A\otimes\A$ (for example, $\Rb_{12}\in
\A\otimes\A\otimes1$, and similarly for $\Rb_{13}$ and
$\Rb_{23}$). Interestingly, the universal $\Rb$-matrix allows one to
construct a map from the tensor square of the algebra $\A$ to itself
\cite{Sklyanin:1988,Kashaev:2004},
\beq
\R:\qquad \A\otimes\A\mapsto \A\otimes\A\,,
\eeq
acting as
\beq
\R:\quad (x \otimes y)\mapsto (x' \otimes y')=\Rb\, (x \otimes y)\,
\Rb^{-1}\,,\qquad x,y\in \A\,.\label{Rmap}
\eeq
It will automatically satisfy Eq.\eqref{SYBE} in virtue of
\eqref{YBE} and commutation relations of the quantum algebra $\A$.
Thus, for any quasitriangular algebra $\A$ the map \eqref{Rmap} 
is a solution the set-theoretic Yang-Baxter equation \eqref{SYBE},
where the algebra $\A$ serves as a set $\X$. At this point it should be
noted that a quasi-triangular 
algebra $\A$ is, of course, not just an abstract structureless  
set (as it is implicitly assumed in the setting of Eq.\eqref{SYBE}), 
but has algebraic
relations between its elements, which must be taken into account 
in order for the Eq.\eqref{SYBE} to hold.
Obviously, this is a slight generalization
of the meaning of Eq.\eqref{SYBE}, but this is a well needed
generalization, since the most important applications of the
Yang-Baxter maps to 
dynamical systems naturally require to equip the set $\X$ with an 
additional structure for purposes of quantization or considerations of
Hamiltonian systems. 
Indeed, we show that the map \eqref{Rmap}
defines  
Hamiltonian evolution equations for a discrete integrable quantum
system in 2D (with discrete space and time) with an algebra of
observables formed by a tensor power of the quantum algebra $\A$. 
The functional Yang-Baxter equation 
\eqref{SYBE} then implies that these evolution equation obey the
quantum version of the consistency-around-a-cube condition
\cite{AdlerBobenkoSuris}. 
Further, the map \eqref{Rmap} possesses a discrete analog of the
zero curvature representation. This allows one to construct a full set
of mutually commuting integrals of motion using the standard approach
of commuting transfer matrices \cite{Bax72} and the
Quantum Inverse Problem Method \cite{Faddeev:1979gh}. The unitary
Heisenberg evolution operator for the system are expressed through the
matrix elements of the universal $\Rb$-matrix. 

Remarkably, all steps of our new approach to the Yang-Baxter maps
admit a meaningful quasiclassical limit. As a result for any
quasitriangular algebra $\A$ one obtains a classical Yang-Baxter map 
which automatically possesses 
properties of a Hamiltonian map, since it
preserves the tensor product structure of Poisson algebras
arising in the quasiclassical limit of the quantum algebra $\A$.
The action functional of the corresponding classical discrete integrable 
system is determined by the quasi-classical asymptotics
of the universal R-matrix.

In this paper we illustrate the above ideas for the case of
the algebra $U_q(sl(2))$, which is the simplest example 
of a quasitriangular Hopf algebra. The quantum case is 
considered in Sect.~2 and the classical one in Sect.~3. The
considerations are completely parallel. For the reader's convenience 
we preserve the same numeration of the corresponding subsections in
both cases. Note the discrete integrable equations arising here for the
algebra $U_q(sl(2))$ are related to the discrete Liouville
equation, various variants of which were 
previously considered in \cite{FT1986,Hirota1987,FV1999,FKV:2001}.

%% The organization of the paper is as
%% follows. In Sect.2 we briefly review
%% basic theory of the algebra $U_q(sl(2))$. Then we calculate
%% the quantum map \eqref{Rmap} explicitly and discuss its key properties.
%% In particular, in Sect.~\ref{evolution} we show that this map describes local 
%% evolution equations for an integrable discrete 2D quantum system with 
%% an algebra of observables formed 
%% by a tensor product of multiple copies of $U_q(sl(2))$, 
%% assigned to edges of a quadrilateral lattice. 
%% In Sect.~\ref{qrmat} we present
%% matrix elements of the quantum $\Rb$-matrix in a factorized form for
%% infinite dimensional representations of $U_q(sl(2))$ in the space of
%% quadratically integrable functions on the real axis, extending the
%% previous results of \cite{BS90, Faddeev:1999, Kashaev:1995,Bytsko2003}.    

%% Sect.3 is devoted to 
%% the quasiclassical limit and the associated classical
%% Yang-Baxter map, arising when $q\to 1$. 
%% It is worth noting that the resulting map automatically possesses 
%% properties of a Hamiltonian map, since it
%% preserves the tensor product structure of Poisson algebras
%% \eqref{poiss}, arising in the quasiclassical limit of $U_q(sl(2))$.
%% Likewise, the generating function of the classical map, which serves as the
%% Lagrangian density of the corresponding evolution system in
%% Sect.~\ref{lagrangian}, is determined by the quasiclassical limit of
%% the universal $\Rb$-matrix.

\section{Quantum Yang-Baxter map for the algebra $U_q(sl(2))$}
\subsection{Algebra $U_q(sl(2))$}
In this subsection we briefly summarise the basic properties of the
algebra $U_q(sl(2))$. A more detailed 
description of this algebra well suited to our purposes can be found in  
\cite{Gould:1995}. The universal enveloping algebra $\A=U_q(sl(2))$ is
generated by 
elements $H,E,F$, satisfying the relations,
\be\label{alg-H}
[H,E]=2\,E\;,\quad [H,F]=-2\,F\;,\quad
[E,F]=(q-q^{-1})\,(q^H-q^{-H})\;.
\ee
Here we use a normalisation, where the elements $E$ and $F$ are multiplied
by the factor $(q-q^{-1})$ with respect to the usual choice. 
Below, it will be convenient to also use the element $K=q^H$, 
for which the commutation relations \eqref{alg-K} become
\be\label{alg-K}
K \,E\,=q^2\,E\,K;,\quad K\,F=q^{-2}\,F\,K\;,\quad
[E,F]=(q-q^{-1})\,(K-K^{-1})\;.
\ee
The quadratic Casimir operator, which commutes with all other elements
of the algebra, has the form 
\be\label{C-def}
C=q^{-1}\, K +q\, K^{-1}+E\,F\,.
\ee

The algebra \eqref{alg-K} 
is a Hopf algebra with the co-multiplication $\Delta$, the co-unit
$\epsilon$, and the antipode $S$. The co-multiplication is a map from
the algebra $\A$ to its tensor square  
\beq
\Delta : \qquad 
\A\mapsto \A\otimes \A, \label{Delta}
\eeq
defined as 
\be
\Delta(K)=K\otimes K\;,\quad
\Delta(E)=E\otimes K+1\otimes E\;,\quad 
\Delta(F)=F\otimes 1+K^{-1}\otimes F\;.
\label{comul}
\ee
The co-unit $\epsilon$ is a map from $\A$ to complex numbers.  
It is defined as 
\beq
\epsilon(K)=1\,,\qquad\epsilon(E)=\epsilon(F)=0,\qquad\label{ep-def}
\eeq
The co-multiplication and co-unit 
define algebra homomorphisms, i.e, $\Delta(ab)=\Delta(a)\,\Delta(b)$
and $\epsilon(ab)=\epsilon(a)\epsilon(b)$.
The antipode $S$ is defined as\footnote{More generally, the defining property 
  of the antipode reads \cite{Gould:1995},
$$
m\circ(S\otimes 1)\circ\Delta(a)=m\circ(1\otimes S^{-1})\circ\Delta(a)=0
$$
where $m$ is multiplication map $m\circ(a\otimes b)=ab$ for algebra
$\A$. For $\A=U_q(sl(2))$ this implies \eqref{S-def}.}
\beq
S(K)=K^{-1}, \qquad S(E)=-E\, K^{-1}, \qquad S(F)=-K\,F\,. \label{S-def}
\eeq
Correspondingly, for the element $H$ one has
\beq
\Delta(H)=H\otimes 1+1\otimes H\,,\qquad 
\epsilon(H)=0\,,\qquad 
S(H)=-H\,,\qquad K=q^H\,.\label{comul-H}
\eeq
Note that the antipode 
is an algebra anti-homomorphism, i.e., $S(ab)=S(b)\,S(a)$.
It satisfies an important properties 
\beq
S(1)=1,\qquad \epsilon\circ S=\epsilon,\qquad (S\otimes S)\circ \Delta =\sigma \circ \Delta \circ S=\Delta'
\circ S\label{S-prop}
\eeq
where $\Delta'$ is another co-multiplication 
obtained from $\Delta$ by interchanging factors in the tensor
product,
\be
\Delta'=\sigma \circ \Delta, \qquad \sigma (x\otimes y)=(y
\otimes x).\label{Dprime}
\ee

\subsection{Universal $\Rb$-matrix}\label{sec:rmat}
The algebra $U_q(sl(2))$ is a {\em quasi-triangular} Hopf algebra. This
means that 
there exists an element $\Rb\in \A\otimes\A$, called the
universal $\Rb$-matrix,  which satisfies the properties
\beq\begin{array}{rcl}\label{Runi-def}
\Delta'(x)\ \Rb&=&\Rb\ \Delta(x)\,,\qquad \forall \ x\in \A,\\[.3cm]
(\Delta\otimes1)\,\Rb&=&\Rb_{13}\,\Rb_{23}\,,\\[.3cm]
(1\otimes\Delta)\,\Rb&=&\Rb_{13}\,\Rb_{12}\,,
\end{array} 
\eeq
where $\Rb_{12}=\Rb\otimes1$, $\Rb_{23}=1\otimes\Rb$ and
$\Rb_{13}=(\sigma\otimes1)\,\Rb_{23}$. Together with \eqref{S-def} and
\eqref{S-prop} this definition implies the quantum Yang-Baxter equation 
\beq
\Rb_{12}\,\Rb_{13}\,\Rb_{23}=\Rb_{23}\,\Rb_{13}\,\Rb_{12}\,,\label{SYBE2}
\eeq
and two simpler relations
\beq
\begin{array} {rcccl}
(\epsilon \otimes 1)\,\Rb&=&(1\otimes
  \epsilon)\,\Rb&=&(1\otimes1),\\[.3cm]
(S\otimes 1)\,\Rb&=&(1\otimes
  S^{-1})\,\Rb&=&\Rb^{-1}.\\[.3cm]
\end{array}\label{simple}
\eeq
Note, that the universal $\Rb$-matrix is not
unique as an element of $\A\otimes\A$. 
%(for instance, if $\Rb_{12}$ satisfies \eqref{Runi-def}, then so does
%$\Rb_{21}^{-1}$) 
However, if one 
assumes, that $\Rb\in{\cal B}_-\otimes{\cal B}_+$, where ${\cal
  B}_+$ and ${\cal B}_-$ are Borel subalgebras of $\A$, 
generated by the elements $(H,F)$ and $(H,E)$ respectively,  then 
the universal $\Rb$-matrix is uniquely defined by the following formal
series\cite{Dri87}  
\be
\Rb\,=\,q^{{\frac{H\otimes H}{2}}}\  
\prod_{k=0}^\infty \,\left(1-q^{2k+1}\,E\otimes F\right)\,.\qquad
\label{drinf}
\ee
\subsection{Quantum Yang-Baxter map}
Let $X=\{K,E,F\}$ denotes the set of generating elements of the algebra $\A$
and $X_{1,2}$ denote these sets in the corresponding components of
the tensor product $\A\otimes\A$,
\beq
X_1=X\otimes 1\,,\qquad X_2=1\otimes X\,,\qquad X=\{K,E,F\}\,.
\eeq
Let us now explicitly calculate the map \eqref{Rmap}. Evidently, it is
completely determined by its action on the generating elements of the
tensor product, 
\beq
\R:\qquad (X_1,X_2)\mapsto (X_1',X_2'), \qquad 
X'_i=\Rb \,X_i\,
\Rb^{-1},\qquad i=1,2\,.\label{Rmap2}
\eeq
Using \eqref{drinf} and \eqref{alg-K} one reproduces the result of 
\cite{Kashaev:2004} 
\begin{subequations}\label{map-set12}
\begin{align}
&\left\{\begin{aligned}
K_1'&=K_1\,(1-q^{-1}\,{K_1}^{-1}E_1F_2K_2)\,,\\[.2cm]
E_1'&=E_1\, K_2\,,\\[.2cm]
F_1'& = F_1\,{K_2}^{-1}+F_2
-{K_1}^{-2}\, F_2 \ \big(1-q\,{K_1}^{-1}E_1F_2K_2\big)^{-1}\,,
\end{aligned}\right.\label{map-set1}
\intertext{and}
&\left\{
\begin{aligned}
K_2'&=(1-q^{-1}\,{K_1}^{-1}E_1F_2K_2)^{-1}\,K_2\,,\\[.2cm]
E_2'& = K_1\,E_2+E_1  - E_1\,{K_2}^2 \ \big(1-q\,
{K_1}^{-1}E_1F_2K_2\big)^{-1}\,,\\[.2cm]
F_2'&=K_1^{-1} F_2\,.
\end{aligned}\right.\label{map-set2}
\end{align}
\end{subequations}
Note, that the relatively complicated 
expressions for $F_1'$ and $E_2'$ above can be easily obtained
by combining the remaining four equations in \eqref{map-set12} with the
formulae for the co-multiplications $\Delta$ and $\Delta'$.

For further references present also the inverse map. Solving
\eqref{map-set12} with respect to the ``unprimed'' variables, one obtains 
\begin{subequations}\label{map-inv}
\begin{align}
&\left\{\begin{aligned}
K_1&=K_1'\,\big(1-q^{-1}\,E_1'\,F_2'\big)^{-1}\,,\\[.2cm]
E_1&=E_1'\, (K_2')^{-1}\,\big(1-q^{-1}\,E_1'\,F_2'\big)^{-1},\\[.2cm]
F_1& = \big(F_1'+(K_1')^{-1}\,
F_2'\big)\,\big(1-q^{-1}\,E_1'\,F_2'\big)\, K_2'- 
K_1' \,F_2'\, K_2'\,,
\end{aligned}\right.\label{map-inv1}
\intertext{and}
&\left\{
\begin{aligned}
K_2&=\big(1-q^{-1}\,E_1'\,F_2'\big)\,K_2'\,,\\[.2cm]
E_2&
=\big(E_2'+E_1'\,K_2'\big)\big(1-q^{-1}\,E_1'\,F_2'\big)\,(K_1')^{-1}-
E_1'\,\big(K_1'\,K_2'\big)^{-1}
\,,\\[.2cm]
F_2&=K_1'\,F_2'\, \big(1-q^{-1}\,E_1'\,F_2'\big)^{-1}\,.
\end{aligned}\right.\label{map-inv2}
\end{align}
\end{subequations}

\subsection{Properties of the quantum Yang-Baxter map}\label{qprop}

It is instructive to reformulate the properties of the universal
$\Rb$-matrix, stated above in \eqref{Runi-def}--\eqref{simple}, 
as properties of the quantum map \eqref{map-set12}\footnote{%
As noted above this map was previously obtained
in \cite{Kashaev:2004}, however, its properties, given below
in Eqs.\eqref{RD}--\eqref{RS} are new.}. 
For this purpose is it useful to introduce a ``set-theoretic
multiplication'' $\delta$ which acts on two sets of generating elements
\beq
\delta: \qquad (X_1,X_2)\mapsto X'\,,\label{delta}
\eeq
and write it as
\beq
X'=\delta(X_1,X_2)\,.\label{Xfunc}
\eeq
Explicitly, it is defined as  
\beq
K'=K_1\,K_2,\qquad E'=E_1\,K_2+E_2,\qquad F'=F_1+{K_1}^{-1}\,F_2\,,
\eeq
which are essentially the same formulae as in \eqref{comul}, but their
meaning is rather different. Indeed, it is worth noting that the
direction of the arrow in \eqref{delta} is reversed with respect to
\eqref{Delta} and that the map \eqref{delta} is only defined on two
sets of generating elements. 

\begin{figure}[h]
\centering
\setlength{\unitlength}{.4cm}
\begin{picture}(8,8)(6,-2)
\Thicklines
% delta_{12}
\put(4,2){\circle{2.82}} 
\put(3.1,1.7){\Large$\delta_{12}$}
%X^\prime
\path(0,2)(2.59,2) 
\path(.3,1.7)(0,2)(.3,2.3)
\put(-1,2.5){$X^\prime$}
%X_1
\spline(5,3)(6.5,4)(8,4)
\put(5,3){\circle*{.4}} 
\path(5.2,3.5)(5,3)(5.5,3)
\put(8,4.4){$X_1$}
%X_2
\spline(5,1)(6.5,0)(8,0) 
\path(5.2,.5)(5,1)(5.5,1)
\put(8,-1){$X_2$}
\put(15,2) {$X^\prime=\delta(X_1,X_2)$}
\end{picture}

\begin{picture}(6,9)(6,-2) 
\Thicklines
\path(3,3.6)(3,3)(3.6,3)
\path(3,0.4)(3,1)(3.6,1)
\put(2,2){\circle{2.82}}
\put(3,1){\circle*{.4}}
\put(1,3){\circle*{.4}}
\spline(-3,0)(0,.2)(1,1)
\path(-2.6,-0.4)(-3,0)(-2.6,0.4)
\spline(7,0)(4,.2)(3,1)
\spline(-3,4)(0,3.8)(1,3)
\path(-2.6,3.6)(-3,4)(-2.6,4.4)
\spline(7,4)(4,3.8)(3,3)
\put(-4.9,-.3){\normalsize $X_2'$}
\put(7.5,-.3){\normalsize$X_1$}
\put(-4.9,3.7){\normalsize $X_1'$}
\put(7.5,3.7){\normalsize $X_2$}
\put(1.1,1.7){\Large$\R_{12}$}
\put(14,2){\normalsize $(X_1',X_2')={\mathcal R}(X_1,X_2)$}
\end{picture}
\caption{Graphical representation of the maps \eqref{Xfunc} and
  \eqref{qRfunc}. The first
  arguments are marked with with heavy dots.}  
\label{twomaps}
\end{figure}
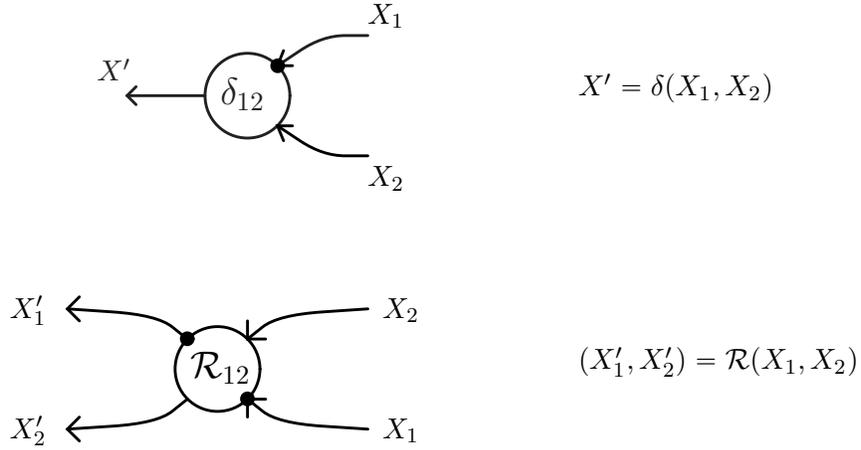

Below, it will also be convenient to write the Yang-Baxter map
\eqref{Rmap2} in a functional form
\beq
(X_1',X_2')=\R(X_1,X_2)\,.\label{qRfunc}
\eeq
Fig.~\ref{twomaps} shows a graphical representation of the maps
\eqref{Xfunc} and \eqref{qRfunc}. Note that these map are not
symmetric upon exchanging their arguments. To avoid an ambiguity
their first arguments on the diagram are marked with heavy dots.

Consider the tensor product $\A\otimes\A\otimes \A$ of three algebras
$\A$ and let $(X_1,X_2,X_3)$ denote three sets of generators in the
corresponding components of the product. Define a functional operator
$\Rcal_{12}$
\begin{subequations}
\beq
\Rcal_{12}(X_1,X_2,X_3):=(X_1',X_2',X_3)=(\R(X_1,X_2),X_3)
\label{Rmap-set}
\eeq
which acts as \eqref{Rmap2} on the first two sets and does not affect
the third one. Similarly, define operators
\beq
\Rcal_{13}=\sigma_{23} \circ \Rcal_{12} \circ \sigma_{23},\qquad
\Rcal_{23}=\sigma_{12} \circ \Rcal_{13} \circ \sigma_{12}.
\eeq
\end{subequations}
At this point it is worth commenting on a relationship between
the operator notations used in Sect.\ref{sec:rmat}, and 
the set-theoretic substitutions $\Rcal_{ij}$ defined here.
Note, that successive 
similarity transformations acting on a list of
generator sets $(X_1,X_2,X_3\ldots)$ 
with quantum operators $\Rb_{ij}$ 
is equivalent to a successive application of the corresponding set-theoretic map
$\Rcal_{ij}$ taken in the {\em reverse}\/ order.
For instance, one can easily verify  
\begin{equation}\begin{array}{rcl}
\boldsymbol{R}_{12}^{} \,\boldsymbol{R}_{13}^{}\, (X_1,X_2,X_3)\,
\boldsymbol{R}_{13}^{-1}\,\boldsymbol{R}_{12}^{-1} &\equiv&
\mathcal{R}_{13}(\mathcal{R}_{12}(X_1,X_2,X_3))\;, 
\\[.4cm]
\boldsymbol{R}_{12}^{} \,\boldsymbol{R}_{13}^{}\,\boldsymbol{R}_{23}
\,(X_1,X_2,X_3) \,\boldsymbol{R}_{23}^{-1}\,
\boldsymbol{R}_{13}^{-1}\,\boldsymbol{R}_{12}^{-1} &\equiv  &
\mathcal{R}_{23}(\mathcal{R}_{13}(\mathcal{R}_{12}(X_1,X_2,X_3)))\;.
\end{array}
\end{equation}
Eq.\eqref{SYBE2} then immediately implies that the set-theoretic 
maps $\mathcal{R}_{ij}$ satisfy the functional Yang-Baxter equation
\begin{equation}
\mathcal{R}_{23}(\mathcal{R}_{13}(\mathcal{R}_{12}(X_1,X_2,X_3)))
=
\mathcal{R}_{12}(\mathcal{R}_{13}(\mathcal{R}_{23}(X_1,X_2,X_3)))\,.
\end{equation}
and, more generally,
\begin{equation}\label{qste}
\mathcal{R}_{23}\circ\mathcal{R}_{13}\circ\mathcal{R}_{12}
=
\mathcal{R}_{12}\circ\mathcal{R}_{13}\circ\mathcal{R}_{23}\,.
\end{equation}

To reformulate the remaining properties of the universal $\Rb$-matrix 
introduce additional notations,
\be
\delta_{12}(X_1,X_2,X_3):=(\delta(X_1,X_2),X_3)\,,
\qquad
\delta_{23}(X_1,X_2,X_3):=(X_1,\delta(X_2,X_3))\,.
\ee
Then Eqs.\eqref{Runi-def} lead to 
\begin{align}
\delta(X_2,X_1)&=\delta\circ \Rcal(X_1,X_2)\,,\label{RD}\\[.2cm]
\Rcal(\delta_{12}(X_1,X_2,X_3))
&=\delta_{12}(\Rcal_{23}(\Rcal_{13}(X_1,X_2,X_3)))\,,\label{DR1} \\[.2cm]
\Rcal(\delta_{23}(X_1,X_2,X_3))
&=\delta_{23}(\Rcal_{12}(\Rcal_{13}(X_1,X_2,X_3)))\,.\label{DR2}
\end{align}
\vspace{.5 cm}
\begin{figure}[h]
%\label{rd-fig}
\centering
\setlength{\unitlength}{.4cm}
\begin{picture}(25,6)(0,-2)
\Thicklines
% delta_{21}
\put(4,2){\circle{2.82}} 
\put(3.1,1.7){\Large$\delta_{21}$}
%\delta_{12}
\put(15,2){\circle{2.82}} 
\put(14.1,1.7){\Large$\delta_{12}$}
%R_{12}
\put(21,2){\circle{2.82}} 
\put(20.1,1.7){\Large$\R_{12}$}
%X^\prime
\path(0,2)(2.59,2) 
\path(.3,1.7)(0,2)(.3,2.3)
\put(-1,2.5){$X^\prime$}
%X_2
\spline(5,3)(6.5,4)(8,4) 
\path(5.2,3.5)(5,3)(5.5,3)
\put(8,4.4){$X_2$}
%X_1
\spline(5,1)(6.5,0)(8,0) 
\put(5,3){\circle*{.4}}
\path(5.2,.5)(5,1)(5.5,1)
\put(8,-1){$X_1$}
%=
\put(8.5,1.8){$=$}
%X^\prime
\path(11,2)(13.59,2) 
\path(11.3,1.7)(11,2)(11.3,2.3)
\put(10,2.5){$X^\prime$}
%X_1^\prime
\spline(16,3)(18,4)(20,3) 
\put(16,3){\circle*{.4}}
\put(20,3){\circle*{.4}}
\path(16.2,3.5)(16,3)(16.5,3)
\put(17.5,4.4){$X_1^\prime$}
%X_2^\prime
\spline(16,1)(18,0)(20,1) 
\path(16.2,.5)(16,1)(16.5,1)
\put(17.5,-1){$X_2^\prime$}
%X_2
\spline(22,3)(23.5,4)(25,4)
\path(22.2,3.5)(22,3)(22.5,3)
\put(25,4.4){$X_2$} 
%X_1
\spline(22,1)(23.5,0)(25,0) 
\put(22,1){\circle*{.4}}
\path(22.2,.5)(22,1)(22.5,1)
\put(25,-1){$X_1$}
\end{picture}

\begin{picture}(29,14)(0,-2)
\Thicklines
% R_{12}
\put(5,7){\circle{2.82}}
\put(4.1,6.7){\Large$\R_{12}$}
% R_{13}
\put(9,3){\circle{2.82}}
\put(8.1,2.7){\Large$\R_{13}$}
% \delta_{23}
\put(1,3){\circle{2.82}}
\put(0.3,2.7){\Large$\delta_{23}$}
% X_2^{\prime\prime}
\path(-3,3)(-0.41,3)
\path(-2.7,2.7)(-3,3)(-2.7,3.3)
\put(-4.5,3.0){$X_2^{\prime\prime}$}
% X_1^{\prime\prime}
\spline(0,9)(2,9)(4,8)
\path(.4,8.6)(0,9)(.4,9.4)
\put(-1,9.5){$X_1^{\prime\prime}$}
% X_2^\prime
\put(2,4){\circle*{.4}}
\put(4,8){\circle*{.4}}
\put(6,6){\circle*{.4}}
\put(8,4){\circle*{.4}}
\put(10,2){\circle*{.4}}
\path(4,6)(2,4)
\path(2.1,4.6)(2,4)(2.7,4)
\put(1.3,5.2){$X_2^\prime$}
% X_1^\prime
\path(6,6)(8,4)
\path(6.1,5.4)(6,6)(6.5,6)
\put(7.5,5.2){$X_1^\prime$}
% X_3^\prime
\spline(2,2)(5,1)(8,2)
\path(2.4,1.5)(2,2)(2.6,2.2)
\put(4.5,-0){$X_3^\prime$}
% X_2
\spline(6,8)(8,9)(10,9)
\put(10,9.5){$X_2$}
\path(6.2,8.5)(6,8)(6.5,8)
% X_3
\spline(10,4)(11.5,5)(12,5)
\put(12,4){$X_3$}
\path(10.2,4.5)(10,4)(10.5,4)
%X_1
\spline(10,2)(11.5,1)(13.5,1)
\put(12,-0){$X_1$}
\path(10.2,1.5)(10,2)(10.5,2)
% = 
\put(14,5){\Large$=$}
% R_{12}
\put(23,5){\circle{2.82}}
\put(22.1,4.7){\Large$\R_{12}$}
% \delta_{23}
\put(28,8){\circle*{.4}}
\put(27,7){\circle{2.82}}
\put(26.1,6.7){\Large$\delta_{23}$}
% X_1^{\prime\prime}
\put(22,6){\circle*{.4}}
\spline(16,9)(18,9)(20,8)(22,6)
\put(15,9.5){$X_1^{\prime\prime}$}
\path(16.3,8.7)(16,9)(16.3,9.3)
% X_2^{\prime\prime}
\spline(17,3)(19.5,3)(22,4)
\path(17.3,2.7)(17,3)(17.3,3.3)
\put(16.5,1.8){$X_2^{\prime\prime}$}
% X_2?
\spline(25.61,6.7)(24.8,6.7)(24,6)
\path(24.6,6)(24,6)(24.1,6.5)
\put(24,7.5){$X_2^\star$}
% X_1
\put(24,4){\circle*{.4}}
\spline(24,4)(27,1)(32.5,1)
\path(24,3.5)(24,4)(24.5,4)
\put(32,0){$X_1$}
% X_3
\spline(28,6)(29.5,5)(32,5)
\path(28.2,5.5)(28,6)(28.5,6)
\put(31,4){$X_3$}
% X_2
\spline(28,8)(29.5,9)(31,9)
\path(28.2,8.5)(28,8)(28.5,8)
\put(30,9.5){$X_2$}
\end{picture}
\caption{Graphical representation of the relations \eqref{RD} and
\eqref{DR2}.}
\label{rd-fig}
\end{figure}
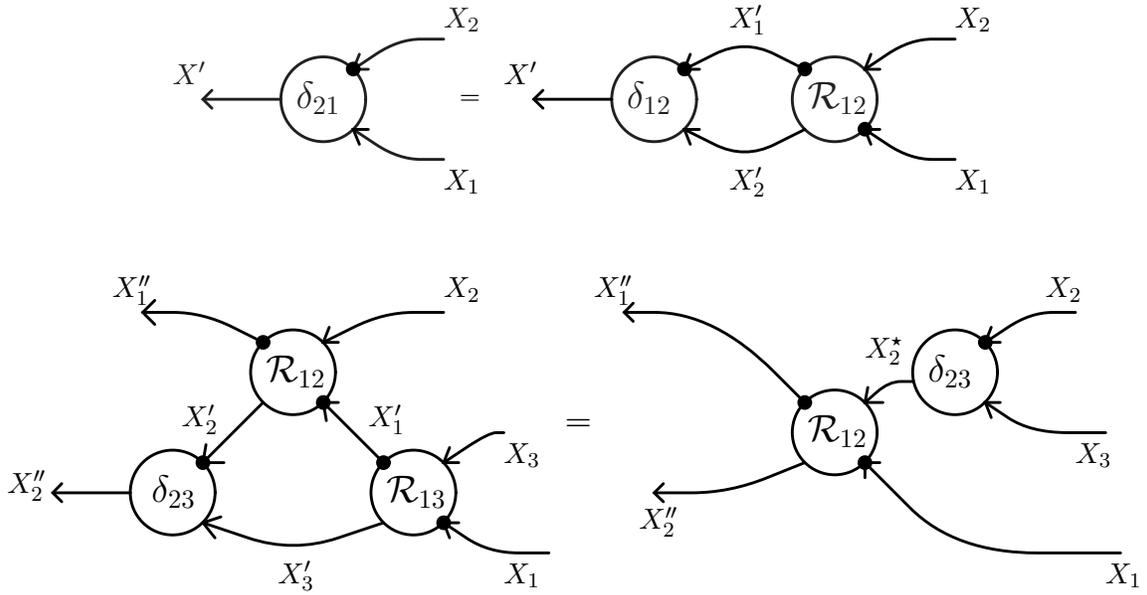

Further, from \eqref{simple} it follows that 
\begin{align}
\R\circ (\epsilon\otimes1)&=(\epsilon\otimes1)\,,\nonumber\\[.2cm]
\R\circ (1\otimes\epsilon)&=(1\otimes\epsilon)\,,\label{eR}\\[.2cm]
\R\circ(S\otimes S)&=(S\otimes S)\circ \R^{-1}\,,\label{RS}
\end{align}
where the co-unit $\epsilon$ and the antipode $S$ are defined in \eqref{S-def}. 
We would like to stress that the above relations \eqref{RD}--\eqref{RS}
are not specific to the algebra $U_q(sl(2))$ and must hold for any 
quasi-triangular Hopf algebra.
In the case of $U_q(sl(2))$ they could be verified using 
the explicit form of the map $\R$ and its inverse given in
\eqref{map-set12} and \eqref{map-inv}. The equations \eqref{RD} and
\eqref{DR2} are illustrated in Fig.~\ref{rd-fig}

\subsection{$\Rb$-matrix form of the $U_q(sl(2))$ defining relations}
\label{R_mat_form} 
As remarked before the universal $\Rb$-matrix is not uniquely defined by
the relations  \eqref{Runi-def}. One can easily to check, that if
$\Rb_{12}\in {\cal B}_-\otimes {\cal B}_+$, \ given by \eqref{drinf},  
satisfies  \eqref{Runi-def}, then so does the element 
\beq\label{Rstar}
\Rb^{*}_{12}=\Rb_{21}^{-1}\,\in {\cal B}_+\otimes {\cal B}_-\,,
\eeq
which does not coincide with $\Rb_{12}$ (this is why the algebra $U_q(sl(2))$
is called a {\em quasi-triangular} Hopf algebra). 
In particular, the
new element $\Rb^{*}$ satisfies the Yang-Baxter equation 
\begin{subequations}
\beq
\Rb_{12}^{*}\ \Rb_{13}^{*}\ \Rb_{23}^{*}=
\Rb_{23}^{*}\ \Rb_{13}^{*}\ \Rb_{12}^{*}\,,\label{SYBE3}
\eeq
which is a simple corollary of \eqref{SYBE2} and
\eqref{Rstar}. Similarly, one can derive ``mixed'' relations
\begin{eqnarray}
\Rb_{12}^{*}\ \Rb_{13}\ \Rb_{23}&=&
\Rb_{23}\ \Rb_{13}\ \Rb_{12}^{*}\ ,\label{SYBE4a}\\[.3cm]
\Rb_{12}\ \Rb_{13}\ \Rb_{23}^{*}&=&
\Rb_{23}^{*}\ \Rb_{13}\ \Rb_{12}\ ,\\[.3cm]
\Rb_{12}\ \Rb_{13}^{*}\ \Rb_{23}^{*}&=&
\Rb_{23}^{*}\ \Rb_{13}^{*}\ \Rb_{12}\ ,\\[.3cm]
\Rb_{12}^{*}\ \Rb_{13}^{*}\ \Rb_{23}&=&
\Rb_{23}\ \Rb_{13}^{*}\ \Rb_{12}^{*}\ .\label{SYBE4d}
\end{eqnarray}\label{SYBE4}
 \end{subequations}

Let $\pi_{\frac{1}{2}}$ denotes the 2-dimensional (spin $\frac{1}{2}$)
representations of the algebra \eqref{alg-H}, 
\beq
\pi_{\frac{1}{2}}(H)=\begin{pmatrix} 1& 0\\ 0& -1
\end{pmatrix}\,,\quad
\pi_{\frac{1}{2}}(E)=(q-q^{-1})\,\begin{pmatrix} 0& 1\\ 0& 0
\end{pmatrix}\,,\quad
\pi_{\frac{1}{2}}(F)=(q-q^{-1})\,\begin{pmatrix} 0& 0\\ 1& 0
\end{pmatrix}\,.\label{2dim}
\eeq
Define the ${\mathbf L}$-operators 
\beq
{\mathbf L}^+=\big(\pi_{\scriptscriptstyle{1/2}}\otimes1\big)
\big(\Rb\big)\,,\qquad 
{\mathbf L}^-=\big(\pi_{\scriptscriptstyle{1/2}}\otimes 1\big)\big(\Rb^*\big)
\,,\label{Lpm-def}
\eeq
by evaluating the universal $\Rb$-matrix in the 2-dimensional
representation in the first space (to be called an auxiliary space).
Using \eqref{drinf} and
\eqref{Rstar}, one gets 
\beq
{\mathbf L}^+=\begin{pmatrix}
q^{\frac{H}{2}} &\quad  q^{\frac{H}{2}}\,F\\[.3cm]
0 &\ q^{-\frac{H}{2}}
\end{pmatrix}\,,\qquad
{\mathbf L}^-=\begin{pmatrix}
q^{-\frac{H}{2}} & 0 \\[.3cm]
-E\,q^{-\frac{H}{2}}&\quad q^{\frac{H}{2}}
\end{pmatrix}\,,
\label{Lpm}
\eeq
which are operator-valued matrices, whose elements belong to
$U_q(sl(2))$ and act in the ``quantum space''. 
Choosing the two-dimensional representation for the
latter 
\beq
R^+=(1\otimes \pi_{\scriptscriptstyle{1/2}})({\bf L^+})= 
(\pi_{\scriptscriptstyle{1/2}} \otimes
\pi_{\scriptscriptstyle{1/2}})(\Rb),\qquad
R^-=(1\otimes \pi_{\scriptscriptstyle{1/2}})({\bf L^-})= 
(\pi_{\scriptscriptstyle{1/2}} \otimes
\pi_{\scriptscriptstyle{1/2}})(\Rb^*)\,,\label{Rpm-def}
\eeq
one gets
the block $R$-matrices 
\beq
R^+=q^{-\frac{1}{2}} \left(\begin{array}{cc:cc}
  q &&& \\[.3cm]
&\quad 1\quad &(q-q^{-1})& \\[.3cm]
\hdashline&\\[-.3cm]
& & 1 &\\[.3cm]
&&&\quad q
\end{array}\right)\,,\quad
R^-=q^{+\frac{1}{2}} \left(\begin{array}{cc:cc}
q^{-1} &&& \\[.3cm]
&1&& \\[.3cm]
\hdashline&\\[-.3cm]
&(q^{-1}-q) & \ 1 &\\[.3cm]
&&&\qquad q^{-1} 
\end{array}\right)\,,
\label{Rpm}
\eeq
where the internal
blocks act in the second space of the product ${\mathbb C}^2\otimes
{\mathbb C}^2$. It is convenient to define the parameter-dependent ${\bf
  L}$-operator 
\beq
{\bf L}(\lambda)=\lambda\, {\bf L}^+ -\lambda^{-1}\,{\bf L}^-\,,\label{Lfull}
\eeq
and the $R$-matrix 
\beq
R(\lambda)=\lambda\, R^+ -\lambda^{-1}\,R^-\,.\label{Rfull}
\eeq
The latter, of course, coincides with the $R$-matrix of the six-vertex
model \cite{Bax82}.
It possesses an important property
\beq
R\big(q^{\frac{1}{2}}\big)=(q-q^{-1})\,P\,,\label{perm}
\eeq
where $P$ is the permutation matrix, interchanging the factors of
the product  ${\mathbb C}^2\otimes {\mathbb C}^2$.
It is useful to define 
\beq
\widecheck R^\pm = R^\pm\, P,\qquad 
\widecheck R(\lambda) = R(\lambda)\, P\,.\label{Rcheck}
\eeq
 Let us now specialize the Yang-Baxter equations \eqref{SYBE2},
and \eqref{SYBE4} by choosing the 2-dimensional
representations \eqref{2dim} in the 
first and second spaces and then using \eqref{Lpm-def} and \eqref{Rpm-def}.
The matrix elements of the ${\bf L}$-operators will then act in the third
space, which now becomes the quantum space. As a result 
one obtains the ``${\bf R}$-matrix form'' \cite{Faddeev:1987ih} of
the commutation relations \eqref{alg-H} 
\beq
\begin{array}{c}\widecheck R^{\pm}\,\big(\,{\bf L}^+\mtimes \,{\bf
    L}^+\,\big)
=\big(\,{\bf
    L}^+\mtimes\,{\bf L}^+\,\big)\,\widecheck R^{\pm}\,,\qquad 
\widecheck R^{\pm}\,\big(\,{\bf L}^-\mtimes\,{\bf L}^-\,\big)=
\big(\,{\bf L}^-\mtimes\,{\bf L}^-\,\big)\,\widecheck R^{\pm}\,,\\[.5cm]
\widecheck R^{+}\,\big(\, {\bf L}^+\mtimes\, {\bf L}^-\,\big)=  
\big(\,{\bf L}^-\mtimes\, {\bf L}^+\,\big)\,\widecheck R^{+}\,,\qquad 
\widecheck R^{-}\,\big(\, {\bf L}^-\mtimes\, {\bf L}^+\,\big)=
\big(\,{\bf L}^+\mtimes\,{\bf L}^-\,\big)\, \widecheck R_{12}^{-}\,.
\end{array}\label{Rform}
\eeq
The symbol $\mtimes$ here denotes the tensor product of two-by-two
matrices ${\bf L}^\pm$, acting in different factors of ${\mathbb
  C}^2\otimes
{\mathbb C}^2$, and operator product for their matrix elements. 
With an account of \eqref{perm} and the definitions 
\eqref{Lfull}, \eqref{Rfull}, 
the above relations can be combined into a
compact form of the parameter-dependent Yang-Baxter equation
\beq
\widecheck 
R(\lambda/\mu)\,\big(\, {\bf L}(\lambda)\mtimes\, {\bf L}(\mu)\,\big)
=\big(\,{\bf L}(\mu)\mtimes\, {\bf L}(\lambda)\,\big)\, \widecheck 
R(\lambda/\mu)\,,
\label{6VLLR}
\eeq
which is well-known in the theory of the six-vertex model (see
\cite{Mangazeev:2014gwa,Mangazeev:2014bqa} for recent advances in this area). 
Finally, note that the co-multiplication $\Delta$, defined in
\eqref{comul}, can be written as
\beq
\Delta({\bf L}^{\pm})={\bf L}^{\pm}\ \otimes\  {\bf L}^{\pm}\label{Lcomul}
\eeq
where the symbol $\otimes $ denotes the matrix product of the two-by-two
matrices \eqref{Lpm} and the tensor product for their operator-valued elements,
belonging to different copies of the algebra $U_q(sl(2))$. There are
two equations in \eqref{Lcomul}, where the superscripts are 
either all pluses or all minuses, simultaneously.

\subsection{Heisenberg-Weyl realisation}
The algebra $U_q(sl(2))$ admits an important homomorphism into the
Heisenberg-Weyl algebra, ${\mathbf W}_q$, generated by two elements
$\uf$ and $\vf$, 
\be
{\mathbf W}_q: \qquad \qquad \uf\, \vf =q^2\, \vf\, \uf\ .\label{weyl}
\ee
Namely, if one sets 
\be
K=\uf,\qquad E=\vf \,(z-q\, \uf),\qquad
F=\vf^{-1} \,(1-q\,z^{-1}\uf^{-1}),
\label{homo}
\ee
where the element $z$ commutes with $\uf$ and $\vf$, then all commutation
relations \eqref{alg-K} will be satisfied in virtue of \eqref{weyl}.
The commuting element $z$ parametrises the Casimir operator \eqref{C-def},
\be
C=z+z^{-1}\;.
\ee
Let us now consider the representation \eqref{homo} as a change of
variable for the set of generators $\{K,E,F\}$ to the set 
$\big\{\uf,\vf,z\big\}$ 
and rewrite the map \eqref{map-set12} in the new variables. Assume the
same meaning to the subscripts $1,2$ as in \eqref{map-set12} (they
distinguish components of the tensor product $\A\otimes \A$), for instance, 
\beq
\uf_1'=\Rcal(\uf_1)=\Rb\,(\uf\otimes 1)\,\Rb^{-1}\,,\qquad
\uf_2'=\Rcal(\uf_2)=\Rb\,(1\otimes\uf)\,\Rb^{-1}\,, \label{uvmap-def}
\eeq
and similarly for $\vf_{1,2}$ and $z_{1,2}$.
From \eqref{map-set12} it follows then
\begin{equation}
\label{uvmap}
\left\{
\begin{array}{ll}
\R(\uf_1)=\uf_1\, \gf\,,\quad,&\R(\uf_2)=\gf^{-1}\uf_2\,,\quad 
\\[.3cm]
\R(\vf_1)=\ds\left(\vf_2^{-1}+(\vf_1^{-1}-\frac{q}{z_2}\,\vf_2^{-1})\,
\uf_2^{-1}\right)^{-1}\,,\quad
&\R(\vf_2) = \ds\frac{z_1}{z_2}\vf_1^{} +
(\vf_2^{}-\frac{q}{z_2}\,\vf_1^{})\,\uf_1^{}\,,\\[.5cm]
\R(z_1)=z_1,\quad &\R(z_2)=z_2\,.
\end{array}\right.
\end{equation}
where we have used the notation
\beq
\gf=1-q^{-1}\uf_1^{-1}\vf_1^{}\,(z_1^{}-q\uf_1^{})\,\vf_2^{-1}
\,(\uf_2^{}-qz_2^{-1})\;.
\eeq
Note that the commuting elements $z_1$ and $z_2$, obviously, remain
invariant. They enter the map \eqref{uvmap} as (spectral) parameters.  
Similarly for the inverse map, one obtains
\begin{equation}
\label{uvmap-inv}
\left\{
\begin{array}{ll}
\R^{-1}(\uf_1)=\uf_1\, \widetilde{\gf}^{-1}\,,
&\R^{-1}(\uf_2)=\widetilde{\gf}\, \uf_2\,,\\[.3cm]
\R^{-1}(\vf_1)=\ds\left(\frac{z_1}{z_2}
\,\vf_2^{-1}+(\vf_1^{-1}-\frac{z_1}{q}\,\vf_2^{-1})\,\uf_2^{}\right)^{-1}, 

&\R^{-1}(\vf_2)=
\ds\vf_1^{}+(\vf_2^{}-\frac{z_1}{q}\,\vf_1^{})\,\uf_1^{-1},
\\[.5cm]
\R^{-1}(z_1)=z_1,&\R^{-1}(z_2)=z_2\,.
\end{array}\right.
\end{equation}
where
\be
\widetilde{\gf}=1-q^{-1}\vf_1^{}(z_1-q\uf_1^{})\,\vf_2^{-1}\,(1-qz_2^{-1}
\uf_2^{-1})\;.
\ee
\begin{figure}[h]
\centering
\setlength{\unitlength}{1cm}
\begin{picture}(7,6)(0,0) 
\Thicklines 
\put(1.5,3){\circle*{.1}}
\put(4.5,3){\circle*{.1}}
\put(3.5,5){\circle*{.1}}
\put(2.5,1){\circle*{.1}}
\put(.5,2.9){\normalsize \bf (a)}
%\path(1.5,0)(3,2)(2.5,4)(1,2)(1.5,0)
\path(2.5,1)(4.5,3)(3.5,5)(1.5,3)(2.5,1)
\put(1.6,1.6){\normalsize $\A_1^{}$}
\put(3.5,1.6){\normalsize $\A_2^{}$}
\put(2.1,4.2){\normalsize $\A_2'$}
\put(4.1,4.2){\normalsize $\A_1'$}
\put(2.8,2.8){\Large$\R$}
%\put(-3,2.9){\Large$\R(\A_1,\A_2|\A_1',\A_2')=$}
\end{picture}
\begin{picture}(6,6)(0,0) 
\Thicklines 
\put(.0,2.9){\normalsize \bf (b)}
\put(1.5,3){\circle*{.1}}
\put(4.5,3){\circle*{.1}}
\put(3.5,5){\circle*{.1}}
\put(2.5,1){\circle*{.1}}
%\path(1.5,0)(3,2)(2.5,4)(1,2)(1.5,0)
\put(.9,2.9){\normalsize{$Q_0$}} 
\put(3.4,5.2){\normalsize{$Q'_1$}} 
\put(2.4,0.55){\normalsize{$Q_1$}} 
\put(4.7,2.9){\normalsize{$Q_2$}} 
\path(2.5,1)(4.5,3)(3.5,5)(1.5,3)(2.5,1)
\put(1.5,1.6){\normalsize ${\bf L}_1^{-}$}
\put(3.5,1.6){\normalsize ${\bf L}_2^{+}$}
\put(2.1,4.2){\normalsize $\widetilde{\bf L}_2^+$}
\put(4.1,4.2){\normalsize $\widetilde{\bf L}_1^-$}
\put(2.8,2.8){\Large$\R$}
%\put(-3,2.9){\Large$\R(\A_1,\A_2|\A_1',\A_2')=$}
\end{picture}

\caption{{\bf (a)} graphical illustration of the map
  $\R$, defined by \eqref{Amap}, 
showing the assignment of the algebras
  $\A_1,\A_2,\A_1',\A_2'$ to the edges of a quadrilateral. \ \ 
{\bf (b)} graphical illustration for the zero curvature relation
\eqref{ZCRb}. The ${\bf L}$-operators play a role of edge transition matrices 
for an auxiliary linear problem.} 
\label{quad1}
\end{figure}
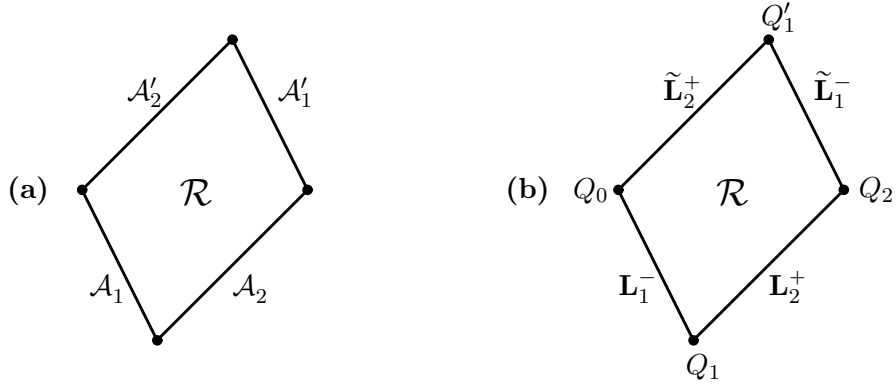
\subsection{Discrete quantum evolution system\label{evolution}}
Here we introduce a $(1+1)$-dimensional integrable quantum evolution
system with discrete space and time. In doing this we  
follow the scheme suggested in \cite{FV:1993,FV:1994} and further
developed in \cite{Bobenko-pendulum,Bazhanov:1995zg,Bazhanov:2008np}.
Remind that the quantum
Yang-Baxter map \eqref{Rmap2} acts on a tensor square of
the algebra $\A$,
\beq
\R:\quad (\A_1 \otimes \A_2)\mapsto (\A_1' \otimes \A_2')=\Rb\, (\A_1 \otimes \A_2)\,
\Rb^{-1}\,.\label{Amap}
\eeq
It can be represented graphically as in Fig.\ref{quad1}(a).
We use this map to define a discrete quantum evolution
system, where the algebra of observables 
\beq
{\mathcal
  O}=\A_1\otimes\A_2\otimes\cdots\otimes\A_{2N-1}\otimes\A_{2N},\qquad
N\ge 1\,, \label{observ}
\eeq
is formed by a tensor power of the algebra $\A$ (all factors 
here are just independent copies of the algebra $\A$; the indices
indicate positions of the corresponding factors in the
product). Let 
\beq
\widecheck\R_{ij}=\sigma_{ij}\circ \R_{ij} 
\eeq
denotes a composition of the map \eqref{Amap}, acting on the $i$-th
and $j$-th factors of the product \eqref{observ}, followed by the
permutation operator $\sigma_{ij}$, swapping these factors with each other. 
With this notation we can define the map
\beq
{\mathcal U}={\mathcal S}\circ \Big(\widecheck\R_{12}\circ
\widecheck\R_{34}\circ\cdots\circ\widecheck\R_{(2N-1),2N}\Big)\,,
\label{evolmap}
\eeq
where the operator ${\mathcal S}$ cyclically shifts the factors in 
the product \eqref{observ},
\beq
{\mathcal S}: \qquad \A_n\to \A_{n+1}, \qquad \A_{2N+1}\equiv\A_1\,.\label{Sdef}
\eeq 

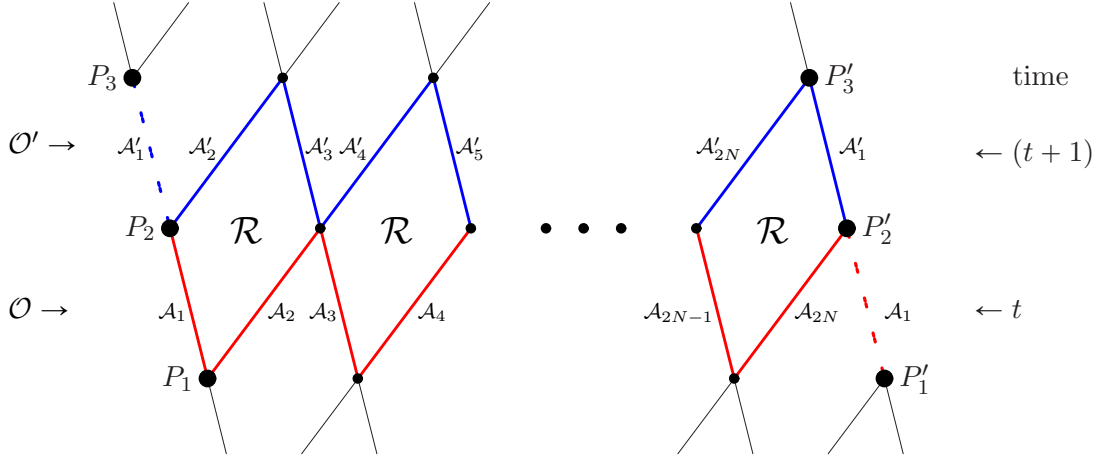
\begin{figure}[h]
\centering
\setlength{\unitlength}{1cm}
\begin{picture}(12,7.)(0,-1.5) 
\put(1,4){
\put(-1.1,-.1){\normalsize{$P_3$}} 
\put(8.7,-.1){\normalsize{$P'_3$}} 
\put(11.2,-.1){\normalsize{time}} 
\put(10.7,-1.1){\normalsize{$\leftarrow(t+1)$}} 
\put(10.7,-3.2){\normalsize{$\leftarrow t$}} 

\path(-.75,1)(-.5,0)(.25,1)
\path(1.25,1)(1.5,0)(2.25,1)
\path(3.25,1)(3.5,0)(4.25,1)
\path(8.25,1)(8.5,0)
\Thicklines
\color{blue}\dashline[5]{.2}[.005](-.5,0)(0,-2)
}
\put(.5,-2){
%\path(-1,2)(-.5,0)(1,2)
\put(.4,1.9){\normalsize{$P_1$}} 

\put(10.2,1.9){\normalsize{$P'_1$}} 
\path(1,2)(1.25,1)
\path(2.25,1)(3,2)
\path(3.,2)(3.25,1)
\path(7.25,1)(8,2)(8.25,1)
\put(2,0){\path(7.25,1)(8,2)(8.25,1)}
}
\Thicklines

\put(.4,1.9){\normalsize{$P_2$}} 

\put(10.2,1.9){\normalsize{$P'_2$}} 
\color{blue}\path(1,2)(2.5,4)(3,2)
\color{red}\path(1,2)(1.5,0)(3,2)
\color{blue}\path(3,2)(4.5,4)(5,2)
\color{red}\path(3,2)(3.5,0)(5,2)
\put(6,2){\circle*{0.1}}
\put(6.5,2){\circle*{0.1}}
\put(7,2){\circle*{0.1}}
\color{red}\path(8,2)(8.5,0)(10,2)
\color{blue}\path(10,2)(9.5,4)(8,2)
%\path(8,2)(8.5,0)(10,2)%(10.5,0)

\color{red}\dashline[5]{.2}[.005](10,2)(10.5,0)
%(9.5,4)(8,2)
%\path(9.5,4)(8,2)(8.5,0)
%
\put(8,2){\circle*{.1}}
\put(5,2){\circle*{.1}}
\put(3,2){\circle*{.1}}

\put(8.5,0){\circle*{.1}}
\put(3.5,0){\circle*{.1}}

\put(4.5,4){\circle*{.1}}
\put(2.5,4){\circle*{.1}}

\put(10,2){\circle*{0.2}}
\put(1,2){\circle*{0.2}}
\put(9.5,4){\circle*{0.2}}
\put(0.5,4){\circle*{0.2}}
\put(10.5,0){\circle*{0.2}}
\put(1.5,0){\circle*{0.2}}

\put(1.8,1.8){\Large$\R$}
\put(3.8,1.8){\Large$\R$}
\put(8.8,1.8){\Large$\R$}
\put(-1.15,.8){\normalsize ${\mathcal O}\to$}
\put(-1.15,3.0){\normalsize ${\mathcal O}'\to$}
\put(0.85,0.8){\scriptsize $\A_1$}
\put(2.3,0.8){\scriptsize $\A_2$}
\put(1.25,3.0){\scriptsize $\A_2'$}
\put(0.3,3.0){\scriptsize $\A_1'$}
\put(2.8,3.0){\scriptsize $\A_3'$}
\put(2.85,0.8){\scriptsize $\A_3$}
\put(4.3,0.8){\scriptsize $\A_4$}
\put(3.25,3.0){\scriptsize $\A_4'$}
\put(4.8,3.0){\scriptsize $\A_5'$}
\put(7.3,0.8){\scriptsize $\A_{2N-1}$}
\put(9.3,0.8){\scriptsize $\A_{2N}$}
\put(8.0,3.0){\scriptsize $\A_{2N}'$}
\put(9.9,3.0){\scriptsize $\A_{1}'$}
%\put(10.5,.8){\scriptsize $\A_{2N+1}=\A_1$}
\put(10.5,.8){\scriptsize $\A_1$}
\end{picture}
\caption{Assignment of dynamical variables to edges of the square
  lattice. Due to periodic boundary conditions the points
  $P_1,P_2,P_3$ on the left side should be identified with their
  images $P'_1,P'_2,P'_3$ on the right side of the lattice.} 
\label{lattice}
\end{figure}

The map \eqref{evolmap} defines an evolution for one step of the
discrete time. By construction it preserves the tensor product
structure of the algebra of observables
\beq
{\mathcal O}'={\mathcal U}\/(\mathcal O)=
\A_1'\otimes\A_2'\otimes\cdots\otimes\A_{2N-1}'\otimes\A_{2N}'
\,.\label{evol}
\eeq
where each algebra $\A_n'$ is isomorphic to the algebra $\A$.
So the evolution map is an automorphism of the algebra of observables,
which preserves its tensor product structure (often referred to as
an ``ultra-local'' structure). For a 
geometric interpretation consider the
square lattice drawn diagonally as in Fig.~\ref{lattice},
with $2N$ sites per row and periodic boundary conditions in the horizontal
(spatial) direction. 
Now assign the factors of the product \eqref{observ} to the set of connected
edges forming a horizontal ``saw'', as shown in
Fig.~\ref{lattice}. In the same way assign the algebras $\A_n'$ in
the product of \eqref{evol} to edges of the saw, shifted by one time
step above from the initial saw. 
Note that these conventions are completely consistent with the graphical
representation of the map \eqref{Amap},
corresponding to one quadrilateral, shown in
Fig.\ref{quad1}(a). 

The elements of the algebra \eqref{observ} constitute the set of dynamical
variables of the system at any fixed value of time.  
With the Heisenberg-Weyl realization \eqref{homo} 
the generating elements of this algebra 
are expressed through the elements $\{\uf_i,\vf_i,z_i\}$, \ where the
index $i =1,2,\ldots 2N$ is considered as the spatial coordinate, 
numerating the factors in \eqref{observ}. 
Combining \eqref{uvmap} with \eqref{evolmap} and
\eqref{evol}, it is easy to see that
\beq
{\mathcal U}(z_{2n+1})=z_{2n-1},\qquad {\mathcal U}(z_{2n})=z_{2n}, 
\qquad n=1,2,\ldots,N\,,
\eeq
where $z_{2N+1}\equiv z_1$. For simplicity, consider the homogeneous
case, when 
\beq
z_{2n-1}\equiv z_1, \qquad z_{2n}\equiv z_2, \qquad \forall n\,,\label{twoval}
\eeq
with arbitrary $z_{1,2}$. Then all $z$'s stay unchanged
under the time evolution, while the remaining variables transforms as 
\begin{equation}
\label{uvevol}
\left\{
\begin{array}{ccl}
\U(\uf_{2n+1})&=&\uf_{2n-1}\, \gf_{n}\,,\\[.3cm]
\U(\vf_{2n+1})&=&\ds\left(\vf_{2n}^{-1}+(\vf_{2n-1}^{-1}-\frac{q}{z_2}\,\vf_{2n}^{-1})\,
\uf_{2n}^{-1}\right)^{-1}\,,\\[.7cm]
\U(\uf_{2n})&=&\gf_{n}^{-1}\uf_{2n}\,,\\[.5cm]
\U(\vf_{2n})& = &\ds\frac{z_1}{z_2}\vf_{2n-1}^{} +
(\vf_{2n}^{}-\frac{q}{z_2}\,\vf_{2n-1}^{})\,\uf_{2n-1}^{}\,,
\end{array}
\right.
\end{equation}
where $n=1,2,\ldots N$, 
\beq
\gf_n=1-q^{-1}\uf_{2n-1}^{-1}\vf_{2n-1}^{}\,(z_1^{}-q\uf_{2n-1}^{})\,\vf_{2n}^{-1}
\,(\uf_{2n}^{}-qz_2^{-1})\;.
\eeq
and the cyclic boundary conditions 
$\uf_{2N+1}=\uf_{1}$ and $\vf_{2N+1}=\vf_{1}$ are implied. With this
realization the algebra of observables at
any fixed moment of the discrete time reduces to the (spatially
localized) tensor product of
the Heisenberg-Weyl algebras \eqref{weyl}, exactly as one would expect
for the equal-time commutation relations of the canonical variables in a
discrete quantum system.

\subsection{Zero curvature representation}\label{sec:ZCR}
Consider again the Yang-Baxter map \eqref{Amap}.
Let ${\bf L}^\pm_1$ and ${\bf L}^\pm_2$ denote ${\bf L}$-operators \eqref{Lpm},
with elements belonging respectively to the first and second algebra of
the tensor product $\A_1\otimes \A_2$, appearing in
\eqref{Amap}. Next, recall the Yang-Baxter equation 
for the universal 
$\Rb$-matrix, given in \eqref{SYBE2}. 
Cyclically shifting its indices $(1,2,3)\to(3,1,2)$ 
and then choosing the 2-dimensional representation \eqref{2dim} 
in the (new) space 3, one obtains 
\beq
{\bf L}_1^+\ {\bf L}_2^+=
\Rb_{12}\ {\bf L}_2^+\ {\bf L}_1^+\ \Rb_{12}^{-1}\,.\qquad \label{LLR3}
\eeq
This equation involves a matrix product of ${\bf L}$'s (as two-by-two matrices)
and a tensor product for their operator-valued matrix elements. Without 
the use of the indices $1,2$, the same equation can be written as 
\beq
{\bf L}^+\otimes\, {\bf L}^+=
\Rb\ \big(\,1\otimes\,{\bf L}^+\,\big)\ (\,{\bf
  L}^+\otimes\,1\,\big)\ \Rb^{-1}\,,\label{LLR4}
\eeq
however we will prefer the indexed form \eqref{LLR3} for greater clarity.  
Further, let 
$\widetilde{\bf L}^\pm_1$ and $\widetilde{\bf L}^\pm_2$ are defined by the
same equations 
\eqref{Lpm}, but with elements belonging, respectively, to 
the algebras $\A_1'$ and $\A_2'$, appearing in \eqref{Amap}.
Then combining \eqref{LLR3} and \eqref{Amap}, one arrives to the equation 
\begin{subequations}
\label{ZCR}
\beq
{\bf L}_1^+\ {\bf L}_2^+=
\widetilde{\bf L}_2^+\ \widetilde{\bf L}_1^+\,,\qquad \label{ZCRa}
\eeq
Similarly, from \eqref{SYBE4a} and \eqref{SYBE4d} one obtains
\beq
{\bf L}_1^-\ {\bf L}_2^+=
\widetilde{\bf L}_2^+\ \widetilde{\bf L}_1^-\,,\qquad
{\bf L}_1^-\ {\bf L}_2^-=
\widetilde{\bf L}_2^-\ \widetilde{\bf L}_1^-\,.\qquad \label{ZCRb}
\eeq
\end{subequations}
For instance, the first equation on the last line reads 
\beq
\begin{array}{c}
(K_1\,K_2)^{{{-\frac{1}{2}}}}
\begin{pmatrix}
K_2 & K_2\, F_2\\[.3cm]
-q\, E_1 \,K_2 & K_1\,\big(\,1-q^{-1}\,K_1^{-1}\,K_2 \,E_1\, F_2\,\big)
\end{pmatrix}
=\hspace{4cm}\\[.9cm]
\hspace{4cm}=(K'_1\,K'_2)^{{{-\frac{1}{2}}}}
\begin{pmatrix}
\big(\,1-q^{-1}\,E'_1\,F'_2\,)\,K'_2 & K'_1\,K'_2\, F'_2\ \\[.3cm]
-q \,E'_1 & K'_1
\end{pmatrix}\,.
\end{array}
\eeq
This equation, as well as the other two matrix equations in \eqref{ZCR} are 
simple corollaries of the explicit expression of the Yang-Baxter map,
given in \eqref{map-set12}. Conversely, the equations \eqref{ZCR} can be
used as an alternative definition of the Yang-Baxter map independent of 
the notion of the universal $\Rb$-matrix.  

Evidently, Eqs.\eqref{ZCR} provide a { ``Zero Curvature
  Representation''} (ZCR) for the map \eqref{map-set12} and for
the local equations of motion of the discrete quantum evolution system
considered in the previous subsection. Indeed, as illustrated in
Fig.\ref{quad1}(b), the two-by-two matrices ${\bf L}$ 
can be viewed as edge transition matrices for an auxiliary discrete
linear problem. Eqs.\eqref{ZCR} then imply that products of 
transition matrices along the paths $Q_0\, Q_1\, Q_2$ and 
$Q_0\, Q_1'\, Q_2$ are exactly the same, i.e., independent of the
choice of path between $Q_0$ and $Q_2$. 

Using the notation \eqref{Lfull} it is convenient to rewrite the ZCR
\eqref{ZCR} in the form
\beq
{\bf L}_1(\lambda)\ {\bf L}_2^+=
\widetilde{\bf L}_2^+\ \widetilde{\bf L}_1(\lambda)\,,\qquad
{\bf L}_1^-\ {\bf L}_2(\lambda)=
\widetilde{\bf L}_2(\lambda)\ \widetilde{\bf L}_1^-\,,\label{ZCR2}
\eeq
where $\lambda$ is arbitrary.

\subsection{Commuting Integrals of Motion}\label{LIM}
The phase space of the quantum evolution system \eqref{evol} 
possesses $4 N$ degrees of freedom ($2N$ coordinates and $2N$
momenta). Let us show that the system has exactly $2 N$ linearly
independent mutually commuting 
integrals of motion (IM). Introduce two transfer matrices 
\begin{subequations}\label{Tdef}
\beq
{\bf T}(\lambda)={\rm Tr}\, \Big( {\bf L}_1(\lambda)\,{\bf L}_2^+\,\,
{\bf L}_3(\lambda)\,{\bf L}_2^+\,\cdots \,{\bf L}_{2N-1}(\lambda)\,{\bf
  L}^+_{2N}\Big)={\rm Tr}\, \prod_{n=1}^N \Big({\bf L}_{2n-1}(\lambda)\,{\bf
  L}^+_{2n}\Big)
\eeq
and
\beq
{\bf \overline T}(\lambda)={\rm Tr}\, \Big( {\bf L}_1^-\,\,{\bf L}_2(\lambda)\,
{\bf L}_3^-\,\,{\bf L}_4(\lambda)\,\cdots \,{\bf L}_{2N-1}^-\,{\bf
  L}_{2N}(\lambda)\Big)={\rm Tr}\, \prod_{n=1}^N \Big({\bf L}_{2n-1}^-\,{\bf
  L}^+_{2n}(\lambda)\Big).
\eeq
\label{Tmats}
\end{subequations}
It follows from \eqref{Rform} and \eqref{6VLLR} that these transfer
matrices form a commuting family
\beq
\big[{\bf T}(\lambda),{\bf T}(\lambda')\big]=
\big[\overline{\bf T}(\lambda),\overline{\bf T}(\lambda')\big]=
\big[{\bf T}(\lambda),\overline{\bf T}(\lambda')\big]=0\,.
\eeq

Further, using the zero curvature representation \eqref{ZCR2} 
one can easily show that the
transfer matrices ${\bf T}(\lambda)$ and $\overline{\bf T}(\lambda)$
are, in fact, time-independent, i.e., they are unchanged under the discrete 
evolution map \eqref{evolmap}, 
\beq
{\bf T}(\lambda)={\mathcal U}\big({\bf T}(\lambda)\big),\qquad
\overline{\bf T}(\lambda)={\mathcal U}\big(\overline
{\bf T}(\lambda)\big)\,.
\eeq

Let $(H_k,E_k,F_k)\in\A_k$ be the set of generators belonging 
to the $k$-th algebra in the tensor product \eqref{observ}. Define
the operators
%\begin{subequations}
\beq
\begin{array}{ll}
{\bf V}^{\,+}_k=q^{-1}\,E_k\ q^{-H_1-H_2-\ldots-H_k}\,,
\qquad
&\overline{\bf V}^+_k=q^{-1}\,E_k\  q^{-H_k-H_{k+1}-\ldots-H_{2N}}\,,
\\[.3cm]
{\bf V}^{\,-}_k=q^{-1}\,F_k\  q^{+H_1+H_2+\ldots+H_k}\,,
\qquad
&\overline{\bf V}^{\,-}_k=q^{-1}\,F_k\  q^{+H_k+H_{k+1}+\ldots+H_{2N}}\,,
\end{array}\label{Vertdef}
\eeq
%\end{subequations}
which obey the following relations
\beq
{\bf V}^{\,\sigma}_k\ {\bf V}^{\,\sigma'}_\ell=
q^{-2\sigma\sigma'}\  {\bf V}^{\,\sigma'}_\ell\ {\bf V}^{\,\sigma}_k\,,\qquad
\overline{\bf V}^{\,\sigma}_k\ \overline{\bf V}^{\,\sigma'}_\ell=
q^{+2\sigma\sigma'}\ \overline{\bf V}^{\,\sigma'}_\ell\ \overline{\bf
  V}^{\,\sigma}_k\,,
\qquad k>\ell\,,
\eeq
\beq
\big[\,{\bf V}^{\,\sigma}_k,\,\overline{\bf
    V}^{\,\sigma}_k\,\big]=0\,,\qquad \big[\,{\bf V}^{\,\sigma}_k,\,\overline{\bf
    V}^{\,\sigma'}_\ell\,\big]=0\,,\qquad k\not=\ell\,,
\eeq
where $\sigma,\sigma'=\pm1$ and 
\beq
\big[\,{\bf V}^{\,+}_k,\,\overline{\bf
    V}^{\,-}_k\,\big]=
\big[\,\overline{\bf V}^{\,+}_k,\,{\bf
    V}^{\,-}_k\,\big]=(q-q^{-1})(q^{H_k}-q^{-H_k})\,.
\eeq

To within simple factors the transfer matrices \eqref{Tmats} are
$N$-th degree polynomials in the variable $\lambda^2$ (or $\lambda^{-2}$),
\beq\label{G-def}
{\bf T}(\lambda)=\lambda^N\ \sum_{n=0}^N \,\lambda^{-2n}\, {\bf G}_n\,,
\qquad
\overline{\bf T}(\lambda)=\lambda^{-N}\ \sum_{n=0}^N \,\lambda^{2n}\,\overline{ {\bf G}}_n
\eeq 
whose coefficients are commuting integrals of motion. 
\beq
\big[{\bf G}_k,{\bf G}_m\big]=
\big[\overline{{\bf G}}_k,{\bf G}_m\big]=
\big[\overline{\bf G}_k,\overline{\bf G}_m\big]\,.
\eeq
From the definition \eqref{Tmats} it is easy to see that there are only 
$2N$ independent coefficients, since two pairs of them coincide
\beq
{\bf G}_0=\overline{\bf G}_0={q}^{\,\bf P}+{q}^{\,-{\bf P}}\,,\qquad
{\bf G}_N=\overline{\bf G}_N\,,\qquad
{\bf P}={\textstyle\frac{1}{2}}\,\sum_{k=1}^{2N} H_k\,.
\eeq
The remaining coefficients can be expressed as ``ordered sums'' of
products of the operators \eqref{Vertdef}, which
are analogous to the ordered integrals that appear in the context of
the continuous conformal field theory \cite{Bazhanov:1996dr} (the so-called
non-local integrals of motion). For instance, the simplest non-trivial
coefficients read
\beq
{\bf G}_1=-\sum_{n=1}^N \Big\{ q^{{\bf P}-H_{2n-1}}
+ q^{{-\bf P}+H_{2n-1}}
%-\sum_{n=1}^N\Big\{
-q^{{\bf P}+1}\sum_{\ell=1}^{2n-2} {\bf V}^{\,-}_\ell\, {\bf
  V}^{\,+}_{2n-1}
-q^{{-\bf P}+1}\sum_{\ell=2 n}^{2N} {\bf
  V}^{\,+}_{2n-1}\,{\bf V}^{\,-}_\ell\Big\}
\eeq   
\beq
\overline{{\bf G}}_1=-\sum_{n=1}^N \Big\{q^{+{\bf P}-H_{2n}}
+ q^{{-\bf P}+H_{2n}}\
-
q^{{\bf P}-1}\sum_{\ell=1}^{2n-1} \overline{{\bf V}}^{\,+}_\ell\, \ \overline{\bf
  V}^{\,-}_{2n}
-q^{{-\bf P}-1}\sum_{\ell=2 n+1}^{2N} \overline{{\bf
  V}}^{\,-}_{2n}\ \overline{\bf V}^{\,+}_\ell\Big\}
\eeq   

Finally note that the transfer matrices \eqref{Tmats} can be viewed as
modified versions of the transfer matrix of the six-vertex model
\cite{Bax82}.

\subsection{Matrix elements of the quantum $\Rb$-matrix\label{qrmat}}

Here we present calculations of the matrix elements of the universal
R-matrix for the representations \eqref{homo}, when the element $\vf$ is
diagonal.  Various parts of these calculations previously appeared in
\cite{Bazhanov:1989nc, Kashaev:1995, Faddeev:1999, Bytsko2003, BMS07a,Kashaev2000}.
The resulting expression \eqref{Rfull} 
is remarkably simple and essentially coincide with that obtained in 
\cite{Bytsko2003}. 
Define new parameters
\be
q=\EXP^{\ii\pi\bb^2}\,,\qquad 
\eta=\frac{1}{2}(\bb+\bb^{-1})\,,\qquad \textrm{Im}(\bb^2)>0\,,\label{eta-def}
\ee
which are simply related to the value of $q<1$. We will also 
assume $|\bb|<1$. Introduce the Heisenberg algebra
\be
\qquad [\xop,\pop]=\frac{\ii}{2\pi}\,. \label{heisenb}
\ee
Below, it will be
convenient to represent the operators entering \eqref{homo} in the form
\be
\uf=\EXP^{-2\pi\bb\, \pop}\;,\qquad \vf=\EXP^{2\pi\bb\,
  (\xop+\ii\beta)}\;,\qquad 
z=-\EXP^{2\pi\ii\bb (\alpha-\beta)}\;,\qquad\label{xp-rep} 
\ee
where $\alpha$ and $\beta$ are arbitrary (complex) parameters. Note,
that the parameter $\beta$ can be eliminated by a trivial redefinition,  
\be
E\to \EXP^{-2\pi \ii \beta}\, E,\qquad
F\to \EXP^{+2\pi \ii \beta}\, F,\qquad 
\alpha\to \alpha+\beta\,,\label{redef}
\ee
however, we prefer to retain it for further convenience.
Next, consider the coordinate representations of the Heisenberg algebra
\eqref{heisenb} in the space of functions $f(\vark)$, 
quadratically integrable
\beq
\int_{i c-\infty}^{ic +\infty} |f(\vark)|^2 
\, d \vark<\infty,\qquad |\textrm{Re}(c)|\le
|\textrm{Re}(\bb)|
\eeq
along any shifted real line  $\textrm{Re}(\vark)\in\mathbb R$ in the
strip  $|\textrm{Im}(\vark)|\le \textrm{Re}(\bb)$,
where the operators $\xop$ and
$\pop$ acts as the multiplication
and differentiation, respectively,
\be 
\xop\,|\vark\rangle=\vark\,|\vark\rangle\,,\qquad
\pop\,|\vark\rangle=-\frac{\ii}{2\pi}\,\frac{d}{ds}\,|\vark\rangle\,.
\label{xdiag}
\ee
The corresponding representation of $U_q(sl(2))$, given by
\eqref{homo}, \eqref{xp-rep} and \eqref{xdiag},  is irreducible
\cite{Bytsko2003} (see also  \cite{DerkachevFaddeev:2014}). 
It will be denoted as
$\pi_{\alpha,\beta}$. It is   
spanned by the vectors $|\vark,\alpha,\beta\rangle$, where\footnote{%
Note, that after the redefinition \eqref{redef} and some other simple
equivalence transformations the representation $\pi_{\alpha,\beta}$ can
be reduced to the representation $\pi_{s}$ defined by Eqs.(2.4) 
of ref.\cite{Bytsko2003} with $s=\ii \alpha$.}  
\beq 
\vf|\vark,\alpha,\beta\rangle  = \EXP^{2\pi\bb
  (\vark+\ii\beta)}|
\vark,\alpha,\beta\rangle,\quad
\uf|\vark,\alpha,\beta\rangle= |\vark-\ii \bb,
\alpha,\beta\rangle
,\quad 
z\,|\vark,\alpha,\beta\rangle=-\EXP^{2\pi\ii\bb(\alpha-\beta)}|\vark,\alpha,\beta\rangle\,.\label{repab}
\eeq 
Further, let
\be|\vark_1,\vark_2\rangle=|\vark_1,\alpha_1,\beta_1\rangle 
\otimes
|\vark_2,\alpha_2,\beta_2\rangle\label{rep12}
\ee
denotes basic vectors in the tensor product
$\pi_{\alpha_1,\beta_1}\otimes\pi_{\alpha_2,\beta_2}$ of two such
representations, where for brevity the dependence on the parameters
$\alpha_{1,2}$ and $\beta_{1,2}$ in the LHS of \eqref{rep12} 
is not explicitly shown. 
It is useful to introduce additional notations
\be
v_i=\EXP^{2\pi\bb (\vark_i+\ii\beta_i)}\;,\quad v_i'=\EXP^{2\pi\bb
  (\vark_i'+\ii\beta_i)}\,, \qquad 
z_i=-\EXP^{2\pi\ii\bb (\alpha_i-\beta_i)}\;\qquad i=1,2.
\ee

Equations \eqref{uvmap-def} and \eqref{uvmap} 
provide recurrence relations for the matrix elements of the $\Rb$-matrix. For
instance, consider the second equation in the left column of  \eqref{uvmap},
namely
\be
\Rb\,\vf_2 = \Big(\frac{z_1}{z_2}\vf_1 + (\vf_2-\frac{q}{z_2}\vf_1)\uf_1\Big)\,\Rb\;.
\ee
If this equation is ``sandwiched'' in between the vectors $\langle s_1,s_2|$ 
and $|s_1',s_2'\rangle$ it gives
\be
\langle s_1^{},s_2^{}|\,\Rb\,|s_1',s_2'\rangle
\ \Big(v_2'-\frac{z_1}{z_2}v_1^{}\Big)  = 
\Big( v_2^{}-\frac{q}{z_2}v_1^{}\Big)\ 
 \langle s_1^{}+\ii\bb,s_2^{}|\,\Rb\,|s_1',s_2'\rangle\;.
\ee
In this way four similar equations in \eqref{uvmap} and
\eqref{uvmap-inv} lead to the following four recurrence relations
\be\label{difference}
\begin{array}{ll}
\ds\frac{\langle s_1^{}+\ii\bb,s_2^{}|\,\Rb\,|s_1',s_2'\rangle}
{\langle s_1^{},s_2^{}|\,\Rb\,|s_1',s_2'\rangle}\;=\;
\frac{\ds z_1\,\bigg(1-\frac{z_2\,v_2'}{z_1\,v_1}\bigg)}{\ds q\,
\bigg(1-\frac{z_2\,v_2}{q\,v_1}\bigg)}\;,  
& \ds\frac{\langle s_1^{},s_2^{}-\ii\bb|\,\Rb\,|s_1',s_2'\rangle}{\langle s_1^{},s_2^{}|\,\Rb\,|s_1',s_2'\rangle}\;=\;
\frac{\ds z_2 \,\bigg(1-\frac{v_2}{v_1'}\bigg)}
{\ds q \,\bigg( 1-\frac{z_2\,v_2}{q\,v_1}\bigg)}\;,\\
[12mm]
\ds\frac{\langle s_1^{},s_2^{}|\,\Rb\,|s_1'+\ii\bb,s_2'\rangle}{\langle s_1^{},s_2^{}|\,\Rb\,|s_1',s_2'\rangle}\;=\;
\frac{\ds\bigg( 1-\frac{v_2}{v_1'}\Bigg)}{\ds {q\,z_1}\,\bigg(1-\frac{v_2'}{q\,z_1\,v_1'}\bigg)}\;,
& \ds\frac{\langle s_1^{},s_2^{}|\,\Rb\,|s_1',s_2'-\ii\bb\rangle}{\langle s_1^{},s_2^{}|\,\Rb\,|s_1',s_2'\rangle}\;=\;
\frac{\ds \bigg(1-\frac{z_2\,v_2'}{z_1\,v_1}\bigg)}{\ds {q\,z_2} 
\bigg(1-\frac{v_2'}{q\,z_1\,v_1'}\bigg)}\;.\\\\
\end{array}
\ee
We assume that the matrix elements 
$\langle
s_1^{},s_2^{}|\,\Rb_{12}\,|s_1',s_2'\rangle$ are analytic in the 
the strip $-\bb\le {\rm Im}s_i,{\rm Im}s'_i \le \bb$. 
Then the above recurrence relations uniquely define these matrix
elements up to an inessential normalization factor,
\be\label{Rfact}
\langle
s_1^{},s_2^{}|\,\Rb_{12}\,|s_1',s_2'\rangle
= 
V_{\beta_1-\alpha_2}(s_2^{}-s_1^{}) V_{\alpha_1-\beta_2}(s_2'-s_1')
\overline{V}_{\alpha_1-\alpha_2}(s_2'-s_1^{})\overline{V}_{\beta_1-\beta_2}(s_2^{}-s_1')\,,
\ee
where 
\beq
V_\alpha(s)=\EXP^{\ii\pi/8-\ii\pi s^2}\varphi(\ii\alpha-s)\;,
\qquad
\overline{V}_\alpha(s)=
\frac{\EXP^{-\ii\pi/8+\ii\pi s^2}}{\varphi(\ii\alpha-\ii\eta-s)}\;.\label{Vdef}
\ee
The function $\varphi(s)$ is the ``non-compact'' quantum dilogarithm
\cite{Faddeev:1999}
\be\label{noncom}
\varphi(s)=\exp\left(\frac{1}{4}\int_{\mathbb{R}+\ii 0}
\frac{\EXP^{-2\ii s
    w}}{\sinh(w\bb)\sinh(w/\bb)}\frac{dw}{w}\right)\;. 
\ee
and the parameter $\eta$ is defined in \eqref{eta-def}. Note that
essentially the same expression \eqref{Rfact} (but in a slightly modified
form) was also obtained in ref.\cite{Bytsko2003}.

The non-compact quantum dilogarithm satisfies the quantum pentagon identity \cite{FKV:2001}
\begin{equation}\label{pentagon}
\varphi(\pop)\varphi(\xop) = \varphi(\xop)\varphi(\pop+\xop)\varphi(\xop)\;,
\end{equation}
where $\xop,\pop$ are the elements of Heisenberg algebra (\ref{heisenb}).

The functions s $V_\alpha(s)$ and $\overline{V}_\alpha(s)$ 
are neither positively defined nor symmetric with respect to the
substitution $s\to-s$.   However, they are Fourier-dual to each other
\be
\int_{\mathbb{R}} 
\EXP^{2\pi\ii sx}\, V_\alpha(s)\,  ds = \overline{V}_\alpha(x)\;.
\ee

The factorized form for the kernel of the $\Rb$-matrix corresponds
to a factorized operator expression acting in the
product of two Heisenberg-Weyl algebra
\eqref{weyl} (but not the full algebra $U_q(sl(2))$),
\be
\Rb_{12} = V_{\beta_1-\alpha_2}(\xop_2-\xop_1)
V_{\alpha_1-\alpha_2}(-\pop_1) V_{\beta_1-\beta_2}(\pop_2)
V_{\alpha_1-\beta_2}(\xop_1-\xop_2) \mathbf{P}_{12}\,, \label{Ropfact}
\ee
where $\mathbf{P}_{12}$ is the permutation operator,
\be
\mathbf{P}_{12}|s_1,\alpha_1,\beta_1\rangle\otimes
|s_2,\alpha_2,\beta_2\rangle = |s_2,\alpha_1,\beta_1\rangle
\otimes|s_1,\alpha_2,\beta_2\rangle,
\ee
which swaps the coordinates $s_1$ and $s_2$, but keep the parameters 
$\alpha_i$ and $\beta_i$ unchanged\footnote{%
It should be note that the factorization of the type \eqref{Rfact},
\eqref{Ropfact} only happens for a specific choice of the basis of
representations of $U_q(sl(2))$. It was first observed in
\cite{Bazhanov:1989nc} for the $\Rb$-matrix of the chiral Potts model,
related to cyclic representations of $U_q(\widehat{sl}(2))$.}.
This $\Rb$-matrix coincides with that for the Dehn twist in quantum Teichm\"uller theory \cite{Kashaev2000}. Explicit correspondence will be established in Appendix.
The elements of the matrix inverse to \eqref{Rfact} are given by
\be\label{Rinverse}
\langle s_1',s_2'|\Rb^{-1}|s_1,s_2\rangle =
\frac{\overline{V}^*_{\alpha_1-\alpha_2}(s_2'-s_1^{})\overline{V}^*_{\beta_1-\beta_2}(s_2^{}-s_1')}{V_{\beta_1-\alpha_2}(s_2^{}-s_1^{})
  V_{\alpha_1-\beta_2}(s_2'-s_1')}\;,
\ee
where
\be
\overline{V}^*_\alpha(s)=\EXP^{\ii\pi/8-\ii\pi
  s^2}\varphi(\ii\eta+\ii\alpha-s)\;.
\ee
This function possesses the property
\be\label{Vinv}
\lim_{\epsilon\to 0}\int ds \overline{V}_{\alpha+\epsilon}(s-u)\overline{V}^*_{\alpha-\epsilon}(s-v)=\delta(u-v)\;,
\ee
where $\alpha$ is purely imaginary and a real positive $\epsilon$ defines
circumventions of the poles. 
Note, that in the case when $\eta$ is real and the spectral parameter $\alpha$ 
is purely imaginary, the superscript
`$*$' can be viewed as the complex conjugation. 
The relation \eqref{Vinv} ensures the fact that the matrices \eqref{Rfact} and
\eqref{Rinverse} are indeed mutually inverse. 

The $\Rb$-matrix \eqref{Rfact} solves the quantum
Yang-Baxter equation\footnote{%
The $\Rb$-matrix \eqref{Rfact} contains two sets of spectral parameters 
$(\alpha_1,\beta_1)$ and $(\alpha_2,\beta_2)$, corresponding to 
the two representations appearing in \eqref{rep12}, 
therefore we write it as 
$\langle s_1^{},s_2^{}|\,\Rb_{12}\,|s_1',s_2'\rangle$. 
In a similar manner the the matrices 
$\langle s_1^{},s_3^{}|\,\Rb_{13}\,|s_1',s_3'\rangle$ 
and $\langle s_2^{},s_3^{}|\,\Rb_{23}\,|s_2',s_3'\rangle$ 
depend on the third set of spectral parameters
belonging to the representation $\pi_{\alpha_3,\beta_3}$ involved in \eqref{ybe-integ}.}
\beq\label{ybe-integ}
\begin{array}{l}
\ds \int_{\mathbb{R}^3} d\vark_1'\,d\vark_2'\,d\vark_3'\ 
\langle \vark_1^{},\vark_2^{}|\,\Rb_{12}\,|\vark_1',\vark_2'\rangle \ 
\langle \vark_1',\vark_3^{}|\,\Rb_{13}\,|\vark_1'',\vark_3'\rangle \ 
\langle \vark_2',\vark_3'|\,\Rb_{23}\,|\vark_2'',\vark_3''\rangle \\
[5mm]
\ds \;\;\;\;\;\;\;\;\;\;\;=\int_{\mathbb{R}^3}
d\vark_1'\,d\vark_2'\,d\vark_3'\
\langle \vark_2^{},\vark_3^{}|\,\Rb_{23}\,|\vark_2',\vark_3'\rangle \ 
\langle \vark_1^{},\vark_3'|\,\Rb_{13}\,|\vark_1',\vark_3''\rangle \ 
\langle \vark_1',\vark_2'|\,\Rb_{12}\,|\vark_1'',\vark_2''\rangle\,.
\end{array}
\eeq
The statement follows from the Yang-Baxter equation \eqref{SYBE2} for
the universal $\Rb$-matrix \eqref{drinf} and the irreducibility of the
representation of $U_q(sl(2))$, given by \eqref{homo} and \eqref{repab}.
It is worth noting that the some solution \eqref{Rfact} can be obtained
as a limiting case of the $\Rb$-matrix of the Faddeev-Volkov model
\cite{FV95,BMS07a}. Thanks to this connection Eq.\eqref{ybe-integ} can be
independently verified using the various star-triangle relations,
associated with the Faddeev-Volkov model and its reductions  
(see Appendix A for further details).  

\tocless\subsubsection{Heisenberg evolution operator}
By construction, it is clear that Eq.\eqref{evol} describes an 
{\em Hamiltonian evolution},\/ in the sense that there exists an invertible
Heisenberg evolution operator $\bf U$,
\beq
{\mathcal U}(\mathcal O)={\bf U}\,{\mathcal O}\, {\bf U}^{-1}
\eeq
which reduces the (discrete time) evolution to 
an internal automorphism of the algebra of observables.
To write this operator explicitly, first remind that we have assumed
that the
elements $z_k$ take only two values, as in \eqref{twoval}. 
Next, consider the representation of algebra \eqref{observ} 
\beq
\pi_{\mathcal O}=\pi_{\alpha_1,\beta_1}\otimes
\pi_{\alpha_2,\beta_2}\otimes\pi_{\alpha_1,\beta_1}\cdots
\otimes\pi_{\alpha_2,\beta_2}
\eeq
given by the product of the representations \eqref{repab} 
with alternating values of the parameters $\alpha$ and $\beta$.
Further, let 
\beq
|\,{\vec s}\,\rangle=|s_1,\alpha_1,\beta_1\rangle\,\otimes\,|s_2,\alpha_2,\beta_2\rangle\,
\otimes\,|s_3,\alpha_1,\beta_1\rangle\,\otimes\ \cdots\ \otimes\,
|s_{2N},\alpha_2,\beta_2\rangle
\eeq
denotes the corresponding basis vectors. Then  
the matrix elements of $\mathbf{U}$ are given by\footnote{%
The matrix elements of $\Rb$ appearing in this formula are given by \eqref{Rfact}
with an appropriate substitution of the variables $s_i$ and $s_i'$.
The parameters $\alpha_{1,2},\beta_{1,2}$ therein remain unchanged.}
\newcommand{\pvec}[1]{\vec{#1}\mkern2mu\vphantom{#1}}
\be
\langle \,\pvec{s}\,|\,\mathbf{U}\,|\,\pvec{s}{\,'}\,\rangle = \prod_{n=1}^N\, \langle
s_{2n-1}^{},s_{2n}^{}|\,\Rb\,|s_{2n+1}',s_{2n}'\rangle 
\ee
The physical regime corresponds to imaginary values of
$\alpha_i,\beta_i$. In this case $\Rb$-matrix is unitary and therefore
the Heisenberg evolution operator is unitary as well.

To write down the matrix elements of the evolution operator for
$T\ge1$ steps of the discrete time introduce $2N(T+1)$ variables
$\{s_{k,t}\}$ with $k=1,2,\ldots,2N$ and $t=0,1,\ldots,T$, such that
\beq
|\pvec{s}\rangle_{(in)}=|\pvec{s},t\rangle\vert_{t=0},\qquad 
|\pvec{s}\rangle_{(out)}=|\pvec{s},t\rangle\vert_{t=T},\qquad |\pvec{s},t\rangle= 
|s_{1,t},\, s_{2,t},\,\ldots,\,s_{2N,t}\rangle
\eeq
describes the initial and final states of the system. Then 
\be\label{partition}
Z={}_{(in)}\langle \pvec{s}|\,\mathbf{U}^T\,|\pvec{s}\rangle_{(out)} = 
\int \Big\{\prod_{t=0}^{T-1}\prod_{n=1}^{N}\, \langle
s_{2n-1,t}^{}\,,s_{2n,t}^{}|\,\Rb\,|s_{2n+1,t+1}\,,s_{2n,t}\rangle\Big\} 
\ \prod_{t=1}^{T-1}\Big(\prod_{k=1}^{2N} ds_{k,t}\Big)
\ee

\section{Quasi-classical limit and classical Yang-Baxter map}

\subsection{Poisson algebra}
Consider the quasi-classical limit
\be
q=\EXP^{\ii\pi\bb^2}\to 1,\qquad \bb\to 0,\label{qclass}
\ee
where the parameter $\bb$ is the same as in \eqref{eta-def}.
In this limit the algebra
\eqref{alg-K} becomes a Poisson algebra (i.e., a commutative algebra
with an associative Poisson bracket $\{\ ,\ \}$, obeying the Leibniz rule). 
Making a substitution 
\be
K\to k,\qquad E\to e,\qquad F\to f,\label{Klim}
\ee
and replacing commutators in \eqref{alg-K} by Poisson brackets,
\beq
[\ ,\ ]\to 2\pi\ii\bb^2\,
\{\ ,\ \}\,,\qquad  \bb\to0\,,
\eeq
one obtains the following Poisson algebra, 
\be {\mathcal P}:\qquad
\{k,e\}=k e,\qquad \{k,f\}=-k f,\qquad 
\{e,f\}=k- k^{-1}\ .
\label{poiss}
\ee
The central element \eqref{C-def} takes the form
\be
C\to \;c=ef+k+k^{-1}\;.\label{C-class}
\ee
The structure of the Hopf algebra in this limit remains intact. In
particular, one can define the co-multiplication, co-unit and antipode,
which, obey the same properties as for the quantum algebra
$U_q(sl(2))$. We will denote them by the symbols $\Dbar$, 
$\ebar$ and $\Sbar$, respectively.
Define the co-multiplication $\Dbar$ as a map 
from the Poisson algebra $\P$ to its tensor square
\beq
\Dbar : \qquad
\P\mapsto \P\otimes \P, \label{Delta-bar}
\eeq
acting as 
\be
\Dbar\,(k)=k\otimes k\;,\quad
\Dbar\,(e)=e\otimes k+1\otimes e\;,\quad 
\Dbar\,(f)=f\otimes 1+k^{-1}\otimes f\;.
\label{comul-bar}
\ee
Similarly to \eqref{ep-def} and \eqref{S-def} 
define the co-unit $\ebar$,
\beq
\ebar\,(k)=1\,,\qquad\ebar\,(e)=\ebar\,(f)=0,\qquad
\eeq
and the antipode $\Sbar$,
\beq
\Sbar\,(k)=k^{-1}, \qquad \Sbar\,(e)=-e\, k^{-1}, 
\qquad \Sbar\,(f)=-k\,f\,. \label{Sbar-def}
\eeq
Note that co-multiplication and co-unit are homomorphisms of the 
Poisson algebra, while the antipode is an antihomomorphisms.
In particular, this means that
\beq
\Dbar\,\big(\big\{a,b\big\}\big)=\big\{\Dbar\,(a),\,\Dbar\,(b)\big\},\qquad 
\Sbar\,\big(\big\{a,b\big\}\big)=-\big\{\Sbar\,(a),\,\Sbar\,(b)\big\}\,.
\eeq
The relations \eqref{S-prop} remain exactly the same as in the quantum
case, 
\beq
\Sbar\,\big(1\big)=1,\qquad \ebar\circ \Sbar=\ebar,\qquad \big(\Sbar\otimes 
\Sbar\big)
\circ \Dbar =\Dbar'\circ S\label{S-prop-classic}
\eeq
where $\Dbar'=\sigma\circ\Dbar$ 
is another co-multiplication 
obtained from \eqref{comul-bar} by interchanging factors in the tensor
product.

\subsection{Quasi-classical expansion of the universal $\Rb$-matrix}
In the limit \eqref{qclass} the universal $\Rb$-matrix \eqref{drinf} becomes
singular. It is not difficult to show that
\be
\Rb=\big(1-e \otimes f\big)^{-\frac{1}{2}}\ \exp\left[\frac{1}{\ii \pi
    \bb^2}\big(\log k\otimes \log k+{\rm Li}_2(e\otimes f)\big)\right]\ 
\Big(1+O(\bb^2)\Big)\,.\label{Rlim}
\ee
where $k,e,f$ are the generators of the Poisson algebra \eqref{poiss}
and ${\rm Li}_2(x)$ is the Euler dilogarithm 
\be
{\rm Li}_2(x)=-\int_0^x \frac{\log(1-t)}{t} \, dt \,.\label{euler-dilog}
\ee

\subsection{Classical Yang-Baxter map}
Even though the quasi-classical limit of the universal $\Rb$-matrix
becomes singular (see \eqref{Rlim} above) its adjoint action on the
elements of the tensor product of two algebras \eqref{alg-K},
appearing in \eqref{Rmap2}, 
is well-defined \cite{Sklyanin:1988}.  Therefore the $q\to 1$ limit of the 
quantum Yang-Baxter map \eqref{Rmap2} 
is well defined.  We will denote it by a special symbol 
\beq\label{Rbar-def}
\Rbar=\lim_{q\to1}\R\,,
\eeq
to distinguish it from its quantum counterpart. 

Let $x=(k,e,f)$ denotes the set of generators
of Poisson algebra $\P$ and $x_{1,2}$ denotes the corresponding sets
in the two components of the tensor product $\P\otimes\P$,
\beq
x_1=x\otimes1,\qquad x_2=1\otimes x, \qquad x=\{k,e,f\}\,.
\eeq
Similarly to \eqref{qRfunc} the classical map
$\Rbar$ is completely defined by is action to two sets of generators
in the tensor square $\P\otimes\P$,
\beq
\Rbar:\qquad(x_1, x_2)\to (x'_1, x'_2)\,,\qquad
(x_1', x_2') ={\Rbar}(x_1, x_2)\,.\label{Ract-bar}
\eeq
The formulae
\eqref{map-set12} lead to 
\begin{subequations}\label{cmap-set12}
\begin{align}
&\left\{\begin{aligned}\label{cmap-set1}
k_1'&=k_1\,\big(1-e_1\,f_2\,k_2/k_1\big)\,,
\\[.2cm]
e_1'&=e_1\, k_2\,, 
\\[.2cm]
f_1'&= \ds{f_1}/{k_2}+f_2
-\ds{f_2}\,{k_1^{-2}\,\big(1-e_1\,f_2\,k_2/k_1\big)^{-1}}\,,\quad\quad 
\end{aligned}\right.
\intertext{and}
&\left\{\begin{aligned}\label{cmap-set2}
k_2'&=\ds{k_2}\,\big({1-e_1\,f_2\,k_2/k_1})^{-1}\,,\\[.2cm]
e_2'&= k_1\,e_2+e_1  - \ds{e_1\,{k_2}^2}\big(
{1-e_1\,f_2\,k_2/k_1}\big)^{-1}\,,\\[.2cm]
f_2'&=f_2/k_1\,.
\end{aligned}\right.
\end{align}
\end{subequations}
We would like to stress that this map preserves the tensor product
structure of two Poisson algebras
\be
{\Rbar}:\qquad {\mathcal P} \otimes {\mathcal P}\to  {\mathcal P}
\otimes {\mathcal 
  P},\label{canonical}
\ee
which means that the ``primed'' quantities $\{k_i', e_i', f'_i\}$,
\ $i=1,2$, have exactly the same Poisson brackets \eqref{poiss}
 as the elements $\{k_i,e_i,f_i\}$. The inverse map reads 
\begin{subequations}\label{cmap-inv}
\begin{align}
&\left\{\begin{aligned}\label{cmap-inv1}
k_1&=k'_1\, \big(1-e_1'\, f_2'\big)^{-1},\\[.2cm]
e_1&=e_1'\,(k_2')^{-1}\,\big(1-e_1'\, f_2'\big)^{-1},\\[.2cm]
f_1&=\big(f_1'+f_2'/k_1'\big)\,\big(1-e_1'\, f_2'\big)\,k_2' 
-k_1'\,f_2'\,k_2',
\end{aligned}\right.
\intertext{and}
&\left\{\begin{aligned}\label{cmap-inv2}
k_2&=k_2' \,\big(1-e_1'\, f_2'\big),\\[.2cm]
e_2&=\big(e_2'+ e_1' k_2'\big)\,\big(1-e_1'\, f_2'\big)/k_1'
  -e_1'/ (k_1'\,k_2')\\[.2cm]
f_2&=f_2'\, k_1'\, \big(1-e_1'\, f_2'\big)^{-1} \ .
\end{aligned}\right.
\end{align}
\end{subequations}
Similarly to the quantum case (see \eqref{RS}) the direct and inverse maps are
related by the antipode \eqref{Sbar-def} 
\be
\Rbar\circ(\Sbar\otimes \Sbar)=(\Sbar\otimes \Sbar)\circ
\Rbar^{-1}\,.
\label{RSbar}
\ee
\subsection{Properties of the classical Yang-Baxter map}
The quasi-triangular structure of the quantum algebra is also  
inherited on the classical level. Therefore, most of the relations of
Sect.~\ref{qprop} remains literally unchanged in the classical case. For
completeness we summarize them below.
Let 
\be
x_i=\{k_i,e_i,f_i\},\qquad i=1,2,3,\ldots 
\ee
be sets of generating elements of different copies of the Poisson
algebra \eqref{poiss}. Introduce the set-theoretic
multiplication $\dbar$ which acts on a two sets generators
\beq
\dbar: \qquad (x_1,x_2)\mapsto x'\,,\label{delta-bar}
\eeq
and write it as
\beq
x'=\dbar(x_1,x_2)\,.
\eeq
Explicitly, one has 
\beq
k'=k_1\,k_2,\qquad e'=e_1\,k_2+e_2,\qquad f'=f_1+{k_1}^{-1}\,f_2\,.
\eeq
which is essentially the same formulae as \eqref{comul-bar}. As in
quantum case, the map \eqref{delta-bar} is only defined on 
two sets of generating element and the direction of the 
arrow in \eqref{delta-bar} is
reversed with respect to \eqref{Delta-bar}. 

Consider the tensor
product $\P\otimes\P\otimes \P$ of three Poisson algebras
\eqref{poiss} 
and let
$(x_1,x_2,x_3)$ denote three sets of generators in the corresponding
components of the product. Define a functional operator
$\Rcal_{12}$
\begin{subequations}
\beq
\Rbar_{12}(x_1,x_2,x_3):=(\Rbar(x_1,x_2),x_3)\label{Rmap-set-bar}
\eeq
which acts as \eqref{Ract-bar} on the first two sets and does not affect
the third one. Similarly, define operators
\beq
\Rbar_{13}=\sigma_{23} \circ \Rbar_{12} \circ \sigma_{23},\qquad
\Rbar_{23}=\sigma_{12} \circ \Rbar_{13} \circ \sigma_{12}.
\eeq
\end{subequations}

In the $q\to 1$ limit Eq.\eqref{qste} simply reduces to the set-theoretic
Yang-Baxter equation for the classical maps ${\Rbar}_{ij}$
\beq
{\Rbar}_{23}\circ{\Rbar}_{13}\circ{\Rbar}_{12}=
{\Rbar}_{12}\circ{\Rbar}_{13}\circ{\Rbar}_{23}\,.\label{cybe}
\eeq
To formulate the remaining properties of the classical map $\Rbar$ 
introduce additional notations,
\be
\dbar_{12}(x_1,x_2,x_3):=(\dbar(x_1,x_2),x_3)\,,
\qquad
\dbar_{23}(x_1,x_2,x_3):=(x_1,\dbar(x_2,x_3))\,.
\ee
Then Eqs.\eqref{Runi-def} lead to 
\begin{align}
\dbar(x_2,x_1)&=\dbar\circ \Rbar(x_1,x_2)\,,\label{RDbar}\\[.2cm]
\Rbar(\dbar_{12}(x_1,x_2,x_3))
&=\dbar_{12}(\Rbar_{23}(\Rbar_{13}(x_1,x_2,x_3)))\,,\label{DR1-classic} \\[.2cm]
\Rbar(\dbar_{23}(x_1,x_2,x_3))
&=\dbar_{23}(\Rbar_{12}(\Rbar_{13}(x_1,x_2,x_3)))\,,\label{DR2-classic}
\end{align}
where the permutation operator $\sigma$ is defined in
\eqref{Dprime}. Further, from \eqref{simple} it follows that 
\begin{align}
\Rbar\circ (\ebar\otimes1)&=(\ebar\otimes1)\,,\nonumber\\[.2cm]
\Rbar\circ (1\otimes\ebar)&=(1\otimes\ebar)\,,\label{eR-classic}\\[.2cm]
\Rbar\circ(\Sbar\otimes \Sbar)&=(\Sbar\otimes \Sbar)\circ
\Rbar^{-1}\,, \label{RS-classic}
\end{align}
where the co-unit $\ebar$ and the antipode $\Sbar$ 
are defined in \eqref{Sbar-def}. 
We would like to stress that the above relations
\eqref{cybe}, \eqref{RDbar}--\eqref{RS-classic}
are not specific to the Poisson algebra \eqref{poiss} and must hold for any 
(quadratic) Poisson algebra which arise in the quasi-classical limit 
of a quasi-triangular Hopf algebra.
In the case under consideration these relations could be verified using 
the explicit form of the map $\R$ and its inverse given in
\eqref{map-set12} and \eqref{map-inv}. 
Note that, contrary to the quantum case, the above relations only involve 
ordinary substitutions for a set of commuting variables, 
since the Poisson algebra $\P$ is a commutative algebra. The fact, that
the map \eqref{canonical} is a ``canonical transformation'' (in the sense
of Hamiltonian dynamics), preserving the Poisson brackets of the
generating elements, does not play any role for the validity of 
\eqref{cybe}-\eqref{RS-classic}.

\subsection{$r$-matrix form of the Poisson algebra}
As noted before, the universal
$\Rb$-matrix \eqref{Rlim} as well as the Yang-Baxter equations
\eqref{SYBE2} and \eqref{SYBE4} becomes singular in the quasiclassical
limit \eqref{qclass}, \eqref{Klim}.
We will return to these singular equations in more details in Sect.\ref{lagrangian}. 

Here we consider the remaining equations in Sect.\ref{R_mat_form},
which all have a smooth limit $q\to 1$. The classical analog of the
${\bf L}$-operators \eqref{Lpm} and \eqref{Lfull} reads
\beq
{\boldsymbol \ell}^+=\begin{pmatrix}
k^{\frac{1}{2}} &\quad k^{\frac{1}{2}} \,f\\[.3cm]
0 &\ k^{-\frac{1}{2}}
\end{pmatrix}\,,\qquad
{\boldsymbol \ell}^-=\begin{pmatrix}
k^{-\frac{1}{2}} & 0 \\[.3cm]
-k^{-\frac{1}{2}}\,e&\quad k^{\frac{1}{2}}
\end{pmatrix}\,,
\label{lpm}
\eeq
and 
\beq
{\boldsymbol \ell}(\lambda)=\lambda\,{\boldsymbol
  \ell}^+-\lambda^{-1}{\boldsymbol \ell}^-\,. 
\eeq
For the $R$-matrix \eqref{Rpm} one obtains
\beq
R^\pm=1\pm\frac{\ii \pi \bb^2}{2}+2 \ii\pi \bb^2\, r^\pm+O(\bb^4),\qquad \bb\to0\,,
\eeq
where
\beq
r^+=\left(\begin{array}{cc:cc}
  0 &&& \\[.3cm]
&\quad -\frac{1}{2} &1 \\[.3cm]
\hdashline&\\[-.3cm]
& & -\frac{1}{2} &\\[.3cm]
&&&\quad 0
\end{array}\right)\,,\qquad
r^-=\left(\begin{array}{cc:cc}
0 &&& \\[.3cm]
&\quad \frac{1}{2}&& \\[.3cm]
\hdashline&\\[-.3cm]
&-1& \ \frac{1}{2} &\\[.3cm]
&&&\qquad 0 
\end{array}\right)\,,
\label{rpm}
\eeq
Similarly to \eqref{Rfull} and \eqref{Rcheck} define
\beq
r(\lambda)=\lambda \, r^+ -\lambda^{-1} \, r^-,\qquad 
\widecheck r(\lambda)=r(\lambda)\,P
\eeq
where $P$ is the permutation operator and $r(\lambda)$ is the
classical $r$-matrix satisfying the classical Yang-Baxter equation \cite{Skl79}
\beq
[r_{12}(\lambda),\,r_{13}(\lambda\mu)]+
[r_{12}(\lambda),\,r_{23}(\mu)]+
[r_{13}(\lambda\mu),\,r_{23}(\mu)]=0\,.
\eeq
The equation \eqref{6VLLR} then
reduces to
\beq
\big\{{\boldsymbol \ell}(\lambda)\,\underset{{}^{\textstyle
    ,}}{\otimes}\,{\boldsymbol \ell}(\mu)\big\} =\big[r(\lambda/\mu),\,
{\boldsymbol \ell}(\lambda)\,{\otimes}\,{\boldsymbol \ell}(\mu)\big]\,,\label{6Vllr}
\eeq
where the symbol $\{\ \underset{{}^{\textstyle,}}{\otimes}\ \}$
denotes Poisson bracket of two factor of the tensor product.

\subsection{Heisenberg-Weyl realisation (canonical variables)}
The three-dimensional Poisson algebra \eqref{poiss} has one central element
\eqref{C-class}. 
Therefore it must essentially reduce to the classical
Heisenberg-Weyl algebra, defined by the bracket 
\be
{W}: \qquad \{u,v\} =u\,v \,.\label{cweyl}
\ee
The required coordinate transformation (cf. \eqref{homo}) is
\be
k=u\;,\quad 
e= v\,(z-u)\;,\quad 
f= v^{-1}(1-z^{-1}u^{-1})\;. \label{chomo}
\ee
The variable $z$ parametrises the central element \eqref{C-class},
\be
c=z+z^{-1}\,,
\ee
and has vanishing Poisson brackets with $u$ and $v$.
In the new variables the map \eqref{cmap-set12} reduces to the classical
analog of \eqref{uvmap}
\be\left\{\begin{array}{ll}
u_1'=u_1 \,g_{cl},\qquad &u_2'=u_2/g_{cl},
\\[.5cm]
v_1'=\displaystyle
v_1v_2u_2 \,\big(v_1\,u_2+(v_2-v_1/z_2)\big)^{-1}\qquad
&v_2'=\displaystyle
%\frac{z_1}{z_2}v_1+(v_2-v_1/z_2)u_1,\\[.5cm]
{z_1}v_1/z_2+(v_2-v_1/z_2)u_1,\\[.5cm]
z_1'=z_1,\qquad &z_2'=z_2
\end{array}\right.\label{uvmap-class}
 \ee
where
\be
g_{cl}=1-\frac{v_1\,(z_1-u_1)(u_2-1/z_2)}{u_1v_2}\;.
\ee
Note that the parameters $z_{1,2}$ remain unchanged under the map. 
Similarly for the inverse map one obtains
\be\left\{\begin{array}{ll}
u_1=u_1'/g_{cl}',\qquad &u_2=u_2'\, g_{cl}',
\\[.5cm]
v_1=\displaystyle
v_1'v_2'\,\big({z_1}v_1'/{z_2}+(v_2'-z_1v_1')\,u_2'\big)^{-1}\qquad
&v_2=\displaystyle
v_1'+(v_2'-z_1v_1')/u_1'\;.
\end{array}\right.\label{uvmap-class-inv}
\ee
where
\be
g_{cl}'=1-\frac{v_1'\,(z_1-u_1')(u_2'-1/z_2)}{v_2'u_2'}\;.
\ee

\subsection{Discrete classical evolution system}

The quantum evolution system considered in Section~\ref{evolution}
admits a straightforward classical limit. Remind, that the classical
Yang-Baxter map \eqref{Ract-bar} acts on the tensor square of the
Poisson algebra \eqref{poiss},
\beq
\overline\R:\quad (\P_1 \otimes \P_2)\mapsto (\P_1' \otimes
\P_2')=\overline\R\big(\P_1 \otimes \P_2\big)\,.\label{Pmap}
\eeq
Correspondingly, the algebra of observables
\eqref{observ} becomes a tensor power Poisson algebra \eqref{poiss},
\beq
{\mathcal
  O}_{{cl}}=\P_1\otimes\P_2\otimes\cdots\otimes\P_{2N-1}\otimes\P_{2N},\qquad
N\ge 1\,.\label{cobserv}
\eeq
Similarly to the quantum case the factors of this product are assigned
to edges of one horizontal saw, corresponding to a fixed moment of the
discrete time, as shown in Fig.~\ref{lattice} (one just need to
replace all $\A$'s there by $\P$'s to get the classical picture).
The classical evolution map is defined as 
\beq
{\mathcal U}_{cl}={\mathcal S}\circ \Big(\widecheck{\overline\R}_{12}\circ
\widecheck{\overline\R}_{34}\circ\cdots\circ\widecheck{\overline\R}_{(2N-1),2N}\Big)\,,
\label{cevolmap}
\eeq
where $\widecheck {\overline\R}_{ij}=\sigma_{ij}\circ {\overline\R}_{ij}$ 
is the map \eqref{Pmap}, 
acting on the $i$-th
and $j$-th factors of the product \eqref{cobserv}, followed by the
permutation operator $\sigma_{ij}$. The operator ${\mathcal S}$ 
cyclically shifts (to the right) the factors in 
the product \eqref{cobserv}, similarly to \eqref{Sdef}. 
The one-step time evolution is defined as 
\beq
{\mathcal O}_{cl}'={\mathcal U}_{cl}\/({\mathcal O}_{cl})=
\P_1'\otimes\P_2'\otimes\cdots\otimes\P_{2N-1}'\otimes\P_{2N}'
\,.\label{cevol}
\eeq

For the realisation \eqref{chomo} each edge will be assigned a triple
of dynamical variables $\{u_i,v_i,z_i\}$, \ $i=1,2,\ldots,2N$. As in
the quantum case we assume the homogeneous arrangement
\eqref{twoval}. Then all $z$'s stay unchanged under the time 
evolution. The remaining variables possess the canonical Poisson
brackets
\beq
\{\,\log u_i\,, \log v_j\,\}=\delta_{ij},\qquad i,j=1,2,\ldots,2N\,. 
\label{canonvar}   
\eeq
which are preserved by the time evolution. 

Taking classical limit in \eqref{uvevol} and introducing discrete time
variable $t$, one immediately obtains the Hamiltonian equations of motion 
\begin{equation}
\label{uvevol-class}
\left\{
\begin{array}{l}
\ds u_{2n+1,t+1}=u_{2n-1,t}\, g_{n,t}\,,\\
[.3cm]
\ds v_{2n+1,t+1}=\ds\left(\frac{1}{v_{2n,t}}+\left(\frac{1}{v_{2n-1,t}}-\frac{1}{z_2v_{2n,t}}\right)\,
\frac{1}{u_{2n,t}}\right)^{-1}\,,\\
[.5cm]
\ds u_{2n,t+1}=u_{2n,t}/g_{n,t}\,,\\
[.3cm]
\ds v_{2n,t+1} = \ds\frac{z_1}{z_2}v_{2n-1,t} +
\left(v_{2n,t}-\frac{1}{z_2}\,v_{2n-1,t}\right)\,u_{2n-1,t}\,,
\end{array}
\right.
\end{equation}
where 
\beq
g_{n,t}=1-\frac{v_{2n-1,t}(z_1-u_{2n-1,t})(u_{2n,t}-1/z_2)}{u_{2n-1,t}v_{2n,t}}\,. 
\eeq

\subsection{Zero curvature representation }\label{sec:zcr}
The quantum zero curvature representation \eqref{ZCR} admits a
straightforward quasi-classical limit. 
Let ${\boldsymbol \ell}_1^{\pm}$ and ${\boldsymbol \ell}_2^{\pm}$
denote the classical ${\boldsymbol \ell}$-operators \eqref{lpm} with
elements  belonging respectively to the first and second Poisson algebra of
the tensor product $\P_1\otimes \P_2$. 
Next, let 
$\widetilde{\boldsymbol \ell}^\pm_1$ and $\widetilde{\boldsymbol
  \ell}^\pm_2$ are defined in a 
similar way, but with elements from the ``transformed'' algebras
$\P_1'$ and $\P_2'$, arising from the map \eqref{Pmap}. Then Eqs.\eqref{ZCR}
lead to the relations 
\beq
{\boldsymbol \ell}_1^+\ {\boldsymbol \ell}_2^+=
\widetilde{\boldsymbol \ell}_2^+\ \widetilde{\boldsymbol \ell}_1^+\,,\qquad 
{\boldsymbol \ell}_1^-\ {\boldsymbol \ell}_2^+=
\widetilde{\boldsymbol \ell}_2^+\ \widetilde{\boldsymbol \ell}_1^-\,,\qquad
{\boldsymbol \ell}_1^-\ {\boldsymbol \ell}_2^-=
\widetilde{\boldsymbol \ell}_2^-\ \widetilde{\boldsymbol \ell}_1^-\,,\qquad \label{zcr}
\eeq
which can be easily verified using explicit formulae for the classical
map \eqref{cmap-set12}. Conversely, the equations \eqref{zcr} can be
used as an alternative definition of the classical Yang-Baxter map
independent of the notion of the universal $\Rb$-matrix for the
quantum algebra.  The geometric interpretation of the relations
\eqref{zcr} as the zero curvature representation remains essentially
the same as in the quantum case, see Fig.\ref{quad1}(b) and the
relevant discussion at the end of Sect.\ref{sec:ZCR}. Finally, note
that Eqs.\eqref{zcr} can be conveniently rewritten as
\beq
{\boldsymbol \ell}_1(\lambda)\ {\boldsymbol \ell}_2^+=
\widetilde{\boldsymbol \ell}_2^+\ \widetilde{\boldsymbol \ell}_1(\lambda)\,,\qquad
{\boldsymbol \ell}_1^-\ {\boldsymbol \ell}_2(\lambda)=
\widetilde{\boldsymbol \ell}_2(\lambda)\ \widetilde{\boldsymbol \ell}_1^-\,,\label{zcr2}
\eeq
where $\lambda$ is arbitrary, similarly to \eqref{ZCR2}.

\subsection{Involutive Integrals of Motion}\label{lim}
The following considerations essentially repeat those for the quantum
case in Sect.\ref{LIM}.    
The phase space of the classical evolution system \eqref{cevol} 
possesses $4 N$ degrees of freedom ($2N$ coordinates and $2N$
momenta). Let us show that the system has exactly $2 N$ integrals of
motion (IM), which are in involution to each other with respect to
the Poisson bracket \eqref{poiss}. The classical analog of the 
transfer matrix in the quantum case 
is the {\em trace of the monodromy matrix} \ for an
auxiliary linear problem. Similarly to \eqref{Tdef} introduce two
such traces
\begin{subequations} \label{tdef}
\beq
{\boldsymbol t}(\lambda)={\rm Tr}\, \Big( {\boldsymbol
  \ell}_1(\lambda)\,{\boldsymbol \ell}_2^+\,\, 
{\boldsymbol \ell}_3(\lambda)\,{\boldsymbol \ell}_2^+\,\cdots
\,{\boldsymbol \ell}_{2N-1}(\lambda)\,{\boldsymbol 
  \ell}^+_{2N}\Big)={\rm Tr}\, \prod_{n=1}^N \Big({\boldsymbol
  \ell}_{2n-1}(\lambda)\,{\boldsymbol\ell}^+_{2n}\Big)\,,
\eeq
and
\beq
{\bf \overline{\boldsymbol t}}(\lambda)={\rm Tr}\, 
\Big( {\boldsymbol \ell}_1^-\,\,{\boldsymbol \ell}_2(\lambda)\,
{\boldsymbol \ell}_3^-\,\,{\boldsymbol \ell}_4(\lambda)\,\cdots \,
{\boldsymbol \ell}_{2N-1}^-\,{\boldsymbol \ell}_{2N}(\lambda)\Big)={\rm Tr}\,
\prod_{n=1}^N \Big({\boldsymbol \ell}_{2n-1}^-\,{\boldsymbol \ell}^+_{2n}(\lambda)\Big)\,.
\eeq
\end{subequations}
It follows from \eqref{6Vllr} that these quantities form an involutive family
\beq
\big\{{\boldsymbol t}(\lambda),{\boldsymbol t}(\lambda')\big\}=
\big\{\overline{\boldsymbol t}(\lambda),\overline{\boldsymbol t}(\lambda')\big\}=
\big\{{\boldsymbol t}(\lambda),\overline{\boldsymbol
  t}(\lambda')\big\}=0\,,
\eeq
for arbitrary values of $\lambda$ and $\lambda'$.

Further, using the zero curvature representation \eqref{ZCR2} 
one can easily show that 
${\boldsymbol t}(\lambda)$ and $\overline{\boldsymbol t}(\lambda)$
are, indeed, time-independent, i.e., they are unchanged under the discrete 
evolution map \eqref{cevolmap}, 
\beq
{\boldsymbol t}(\lambda)={\mathcal U}_{cl}\big({\boldsymbol t}(\lambda)\big),\qquad
\overline{\boldsymbol t}(\lambda)={\mathcal U}_{cl}\big(\overline
{\boldsymbol t}(\lambda)\big)\,.
\eeq
Similarly to \eqref{G-def} in the quantum case, the classical
Integrals of Motion can be defined as coefficients 
of expansions of the polynomials ${\boldsymbol t}(\lambda)$ and 
$\overline{\boldsymbol t}(\lambda)$ in the variable $\lambda^2$.

% Sergey starts

\subsection{The 
Lagrangian equation of motion and the action\label{lagrangian}} 

In view of the quasi-classical correspondence, considered in the
section, the Lagrangian function and the 
action of the classical evolution system \eqref{uvevol-class} is determined
quasi-classical asymptotics of the universal $\Rb$-matrix \eqref{Rlim}.
More specifically, consider the expression for its matrix elements
given by Eq.\eqref{Rfact}. It involves the spectral parameters
$\alpha_{1,2},\,\beta_{1,2}$ and the spin variables $s_{1,2}$ and
$s_{1,2}'$. For a consistent quasi-classical limit these variables should become
infinite, provided that the quantities
\beq\label{newset}
a_i=-2\pi \ii \bb \,\alpha_i,\quad 
b_i=-2\pi \ii \bb \,\beta_i,\quad 
\sigma_i=2\pi \bb \,s_i,\quad 
\sigma'_i=2\pi \bb \,s'_i,\qquad i=1,2,\quad \bb\to0\,,
\eeq
remain finite for $\bb\to 0$. Introduce new functions
\begin{equation}
\lambda_{a}({\sigma})=-\frac{{\sigma}^2}{2}-\textrm{Li}_2(-\EXP^{-{\sigma}-{a}})\,,\qquad 
\overline{\lambda}_{a}({\sigma})=\frac{{\sigma}^2}{2}+\textrm{Li}_2(\EXP^{-{\sigma}-{a}})\;,
\eeq
where the Euler dilogarithm $\textrm{Li}_2(x)$ defined
in \eqref{euler-dilog}. It is useful to note that
\beq
\ds\frac{\partial\lambda_{a}({\sigma})}{\partial
  {\sigma}}=-\log(\EXP^{\sigma}+\EXP^{-{a}}),\qquad
\frac{\partial\overline{\lambda}_{a}({\sigma})}{\partial
  {\sigma}}=\log(\EXP^{\sigma}-\EXP^{-{a}})\;. 
\end{equation}
With these notations one can show that 
\be\label{Rquasi}
\langle s_1,s_2|R|s_1',s_2'\rangle =
\exp\left\{\frac{\ii}{2\pi\bb^2}
\mathcal{L}(\sigma_1,\sigma_2,\sigma_1',\sigma_2')
+O(1)\right\}\quad \textrm{as}\quad \bb\to 0\;,
\ee
where
\begin{equation}\label{Lag}
\mathcal{L}=\lambda_{{b}_1-{a}_2}({\sigma}_2^{}-{\sigma}_1^{})+\lambda_{{a}_1-{b}_2}({\sigma}_2'-{\sigma}_1')+\overline{\lambda}_{{a}_1-{a}_2}({\sigma}_2'-{\sigma}_1^{})+\overline{\lambda}_{{b}_1-{b}_2}({\sigma}_2^{}-{\sigma}_1')\;.
\end{equation}
This formula easily follows from the asymptotics of the functions \eqref{Vdef}
for $\bb\to 0$,
\begin{equation}\label{Vcl}
V_{\frac{\ii a}{2\pi\bb}}\Big(\frac{{\sigma}}{2\pi\bb}\Big) =
\exp\left\{\frac{\ii}{2\pi\bb^2}\lambda_{a}({\sigma})+O(1)\right\}\,,\qquad
\overline{V}_{\frac{\ii{a}}{2\pi\bb}}\Big(\frac{{\sigma}}{2\pi\bb}\Big)=
\exp\left\{\frac{\ii}{2\pi\bb^2}\overline{\lambda}_{a}({\sigma})+O(1)\right\}\,.
\end{equation}

Eqs (\ref{uvevol-class}) are the (first order) Hamiltonian equations
of motion for our dynamical system. To obtain the corresponding (second order) Lagrangian
equations of motion one may consider the quasi-classical asymptotics of
the matrix elements of the evolution operator
\eqref{partition}. Substituting \eqref{Rquasi} therein and calculating the
integral by the saddle point method one gets 
\beq
\log Z =\frac{\ii\,}{2\pi \bb^2}\,{\mathcal A}\big(\{\sigma\}^{(cl)}\big) +O(1) 
\eeq
where ${\mathcal A}\big(\{\sigma\}^{(cl)}\big)$ is the classical action 
of the system 
\beq
\mathcal{A}\big(\{\sigma\}\big)=\sum_{n=1}^N\sum_{t=0}^{T-1} 
\mathcal{L}\bigl(\sigma_{2n-1,t},\,\sigma_{2n,t},\,\sigma_{2n+1,t+1},\,\sigma_{2n,t+1}\bigr)\label{A-class}
\eeq
calculated on the stationary spin configuration $\{\sigma\}^{(cl)}$
defined as
\beq\label{stateq}
\frac{\partial {\mathcal A}\big(\{\sigma\}\big)}{\partial
  \sigma_{k,t}}\Big\vert_{\{\sigma\}=\{\sigma\}^{(cl)}}=0\,,\qquad k=1,\ldots,2N,\quad t=1,\ldots,T-1\,.
\eeq
The boundary spins $\{\sigma_{k,0}\}_{k=1}^{2N}$ and 
$\{\sigma_{k,T}\}_{k=1}^{2N}$ are kept fixed.

The same result can be obtained by the standard method of Hamiltonian mechanics.
Remind that the classical map \eqref{uvmap-class} involves the set of
six independent variables $u_{1,2},v_{1,2},z_{1,2}$. They are simply related
to the variables in \eqref{newset},  
\beq
z_i=-\EXP^{{b}_i-{a}_i},\qquad v_i=e^{{\sigma}_i-{b}_i},\qquad
v_i'=e^{{\sigma}_i'-{b}_i},
\label{newvar}\;.
\eeq
Next, the map \eqref{uvmap-class} is a canonical transformation, preserving the Poisson bracket
\be
\{\,\log u_i,\, \log v_j\}=\delta_{ij},\qquad i,j=1,2
\ee
Its generation function ${\mathcal L}$ is defined by the equations
\be
d {\mathcal L} = \sum_{i=1,2}\big(\,\log u_i' \,d\log v_i' - \log u_i
\,d\log v_i\,\big)\;. \label{diff}
\ee
or
\beq
\frac{\partial {\mathcal L}}{\partial \log v_i}=-\log u_i,\qquad
\frac{\partial {\mathcal L}}{\partial \log v'_i}=\log u'_i,\qquad i=1,2\,.
\eeq
Rewrite (\ref{uvmap-class}) in the form 
\be\label{uvmap-class-1}
\begin{array}{ll}
\ds u_1= z_1^{}\frac{1-z_2v_2'/z_1v_1}{1-z_2v_2/v_1}\;, & 
\ds u_2= z_2^{-1}\frac{1-z_2v_2/v_1}{1-v_2/v_1'},\\[.3cm]
\ds u_1'= z_1^{}\frac{1-v_2'/z_1v_1'}{1-v_2/v_1'}\;, & 
\ds u_2'= z_2^{-1}\frac{1-z_2v_2'/z_1v_1}{1-v_2'/z_1v_1'}.
\end{array}
\ee
Integrating \eqref{diff} and using the variables \eqref{newvar}, one
obtains the same Lagrangian (\ref{Lag}).

\begin{figure}[h]
\centering
\setlength{\unitlength}{1cm}
\begin{picture}(7,6)(0,0)
\Thicklines
\path(2.5,1)(4.5,3)(3.5,5)(1.5,3)(2.5,1)
\put(2.5,1){\circle*{.1}}
\put(4.5,3){\circle*{.1}}
\put(3.5,5){\circle*{.1}}
\put(1.5,3){\circle*{.1}}
\put(0.9,1.6){$v_{2n-1,t}$}
\put(3.4,1.6){$v_{2n,t}$}
\put(1.4,4.2){$v_{2n,t+1}$}
\put(4,4.2){$v_{2n+1,t+1}$}
\end{picture}
\caption{Arrangement of classical variables $v_{2n-1,t}$ and $v_{2n,t}$.} 
\label{lattice3}
\end{figure}
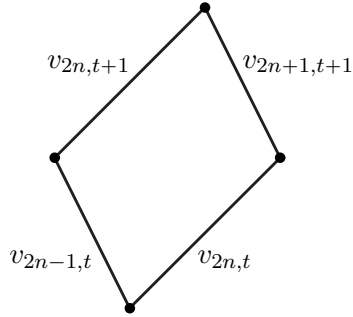

\subsection{Discrete Liouville Equations}
It is instructive to write down the Lagrangian equations of motion
\eqref{stateq} explicitly. The corresponding Hamiltonian
equations \eqref{uvevol-class} are written in terms of the variables
$u_{k,t},v_{k,t}$, \ $k=1,\ldots,2N$, where the 
variables $v_{k,t}$ (playing the role of
coordinates) are related to $\sigma_{k,t}$ in \eqref{A-class} as follows
\beq
v_{2n-1,t}=\EXP^{-b_1+\sigma_{2n-1,t}},\qquad
v_{2n,t}=\EXP^{-b_2+\sigma_{2n,t}},\qquad n=1,\ldots,N,\quad
t=0,\ldots,T\,.\label{v-sigma}
\eeq
They arrangement around an elementary quadrilateral is shown in
Fig.\ref{lattice3}. Rewriting \eqref{uvmap-class} in the form 
\ref{uvmap-class-1} one obtains
\begin{equation}
\label{uvmap-class-again}
\begin{array}{ll}
\ds u_{2n-1,t}=z_1\frac{\ds 1-\frac{z_1}{z_2} \frac{v_{2n,t+1}}{v_{2n-1,t}}}{\ds 1-z_2\frac{v_{2n,t}}{v_{2n-1,t}}}\;, & \ds u_{2n,t}=\frac{1}{z_2}\frac{\ds 1-z_2\frac{v_{2n,t}}{v_{2n-1,t}}}{\ds 1-\frac{v_{2n,t}}{v_{2n+1,t+1}}}\;,\\
[10mm]
\ds u_{2n+1,t+1}=z_1\frac{\ds 1-\frac{1}{z_1} \frac{v_{2n,t+1}}{v_{2n+1,t+1}}}{\ds 1-\frac{v_{2n,t}}{v_{2n+1,t+1}}}\;, & \ds u_{2n,t+1}=\frac{1}{z_2}\frac{\ds 1-\frac{z_2}{z_1}\frac{v_{2n,t+1}}{v_{2n-1,t}}}{\ds 1-\frac{1}{z_1}\frac{v_{2n,t+1}}{v_{2n+1,t+1}}}\;.
\end{array}
\end{equation}
To obtain the Lagrangian equations of motion (\ref{stateq}) one needs
to exclude the variables $u_{k,l}$ (they are regarded as momentum
variables). This is achieved by equating $u$'s defined by the first
line 
of (\ref{uvmap-class-again}) to those from the second line of 
(\ref{uvmap-class-again}) (with an appropriate shift of indices).
\begin{figure}[h]
\centering
\setlength{\unitlength}{1cm}
\begin{picture}(13,8)(0,0)
\Thicklines
\path(2.5,1)(4.5,3)(6.5,5)(5.5,7)(3.5,5)(1.5,3)(2.5,1)
\path(4.5,3)(3.5,5)
\put(2.5,1){\circle*{.1}}
\put(4.5,3){\circle*{.1}}
\put(3.5,5){\circle*{.1}}
\put(1.5,3){\circle*{.1}}
\put(6.5,5){\circle*{.1}}
\put(5.5,7){\circle*{.1}}
\put(5.4,3.6){$v_{r}$}
\put(3.4,1.6){$v_{d}$}
\put(2,4.2){$v_{\ell}$}
\put(4,4.2){$v$}
\put(4,6.2){$v_u$}
\path(10.5,1)(12.5,3)(11.5,5)(10.5,7)(8.5,5)(9.5,3)(10.5,1)
\path(9.5,3)(11.5,5)
\put(10.5,1){\circle*{.1}}
\put(12.5,3){\circle*{.1}}
\put(11.5,5){\circle*{.1}}
\put(10.5,7){\circle*{.1}}
\put(8.5,5){\circle*{.1}}
\put(9.5,3){\circle*{.1}}
\put(10.4,3.6){$v$}
\put(12,4.2){$v_r$}
\put(9,4.2){$v_{\ell}$}
\put(11,6.2){$v_u$}
\put(10,2.2){$v_d$}
%                                                                           
%\draw (2.8,2.8) node[right] {$\pt$};
%
%
\end{picture}
\caption{Arrangement of 
variables for the Lagrangian equations of motion.} 
\label{star}
\end{figure}
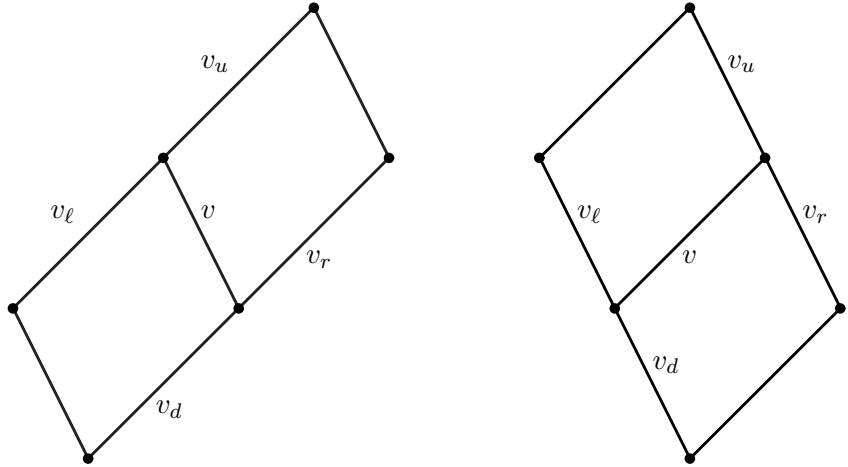
Consider a geometric structure of the resulting equations.
Let $v=v_{k,t}$ be one of the $v$-variables 
for some fixed values of $k,l$. 
Define four adjacent variables
\beq
v_u=v_{k+1,t+1},\quad  
v_d=v_{k-1,t-1},\quad  
v_\ell=v_{k-1,t},\quad  
v_r=v_{k+1,t},\quad  v=v_{k,t}\,,\label{up-down}
\eeq
shifted one site ``up'', ``down'', ``left'' or ``right'' with respect
to $v$, as shown in Fig.\ref{star}. The equations \eqref{stateq} has
different form depending on whether the spatial coordinate of the
central site $(k,t)$ is odd or even. For an odd $k=2n-1$ (left part of Fig.\ref{star}) one obtains
\begin{subequations}\label{odd-even}
\begin{equation}\label{odd}
{\ds \Big(1-\frac{v_\ell}{z_1\,v}\Big) \ \Big(1-\frac{z_2\,v_r}{v}\Big)}=
{\ds \Big(1-\frac{z_2\,v_u}{z_1\,v}\Big)\ \Big(1-\frac{v_d}{v}\Big)^{\phantom{\Big|}}} \,,
\end{equation}
and for an even $k=2n$ (right part of Fig.\ref{star}),
\begin{equation}\label{even}
{\ds \Big(1-\frac{v}{z_1\,v_r}\Big) \Big(1-\frac{z_2\,v}{v_\ell}\Big)}=
{\ds \Big(1-\frac{z_2\,v}{z_1\,v_d}\Big) \Big(1-\frac{v}{v_u}\Big)}\,.
\end{equation}
\end{subequations}
\tocless\subsubsection{General solution of the equation of motion}
\begin{figure}[h]
\centering
\setlength{\unitlength}{1cm}
\begin{picture}(7,6)(0,0)
\Thicklines
\path(2.5,1)(4.5,3)(3.5,5)(1.5,3)(2.5,1)
\put(2.5,1){\circle*{.1}}
\put(4.5,3){\circle*{.1}}
\put(3.5,5){\circle*{.1}}
\put(1.5,3){\circle*{.1}}
\put(1.5,1.6){$v_{1,{\bf x}}$}
\put(3.4,1.6){$v_{2,{\bf x}}$}
\put(1.4,4.2){$v_{2,{\bf x}+{\bf e}_2}$}
\put(4,4.2){$v_{1,{\bf x}+{\bf e}_1}$}
\put(2.8,2.8){${\bf x}$}
\put(2.3,0.6){$\tau_{\bf x}$}
\put(4.7,2.9){$\tau_{{\bf x} + {\bf e}_2}$}
\put(0.5,2.9){$\tau_{{\bf x}+{\bf e}_1}$}
\put(3.5,5.2){$\tau_{{\bf x}+{\bf e}_1+{\bf e}_2}$}
\end{picture}
\caption{Light cone coordinates, ${\bf x}=x_1{\bf e}_1+x_2{\bf e}_2$.} 
\label{quad2}
\end{figure}
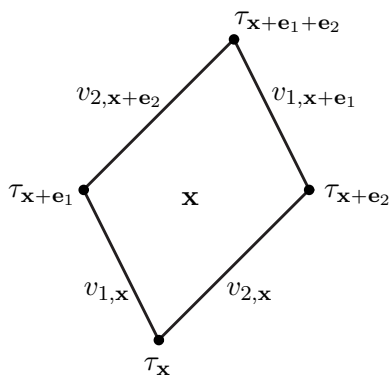

Below we present a general solution of both the Lagrangian and
Hamiltonian equations of motion. 
It will be convenient to introduce the ``light cone'' coordinates: 
\beq
\qquad {\bf x}=(n,t-n),\qquad {\bf e}_1=(1,0), \qquad {\bf e}_2=(0,1)\;.
\eeq
Conversely, 
\beq
n=x_1,\qquad t=x_1+x_2,\qquad {\bf x}=(x_1,x_2)\,.
\eeq
Moreover, we will use the variables $u_{1,2}({\bf x})$  and $v_{1,2}({\bf x})$
instead of $u_{k,t}$ and $v_{k,t}$, 
\beq
{u}_{1}({\bf x})=u_{2n-1,t}, \qquad u_{2}({\bf x})=u_{2n,t},\qquad
{v}_{1}({\bf x})=v_{2n-1,t}, \qquad v_{2}({\bf x})=v_{2n,t}\;.\label{uv-change}
\eeq
The Hamiltonian equations of motion in these notations are easily obtained from
\eqref{uvmap-class-1} by substituting the variables $\{v_1,v_2,v'_1,v'_2\}$
therein with $\{v_1({\bf x}),v_2({\bf x}),v_1({\bf x}+{\bf
  e}_1),v_2({\bf x}+{\bf e}_1)\}$ and similarly for
$\{u_1,u_2,u'_1,u'_2\}$. Their arrangement on lattice is shown in
Fig.~\ref{quad2} (cf. Fig.~\ref{lattice3}).   
The resulting equations imply a simple relation 
\begin{equation}\label{uuuu}
u_1({\bf x}) u_2({\bf x}) = u_1({\bf x}+{\bf e}_1) u_2({\bf x}+{\bf e}_2)\;.
\end{equation}
which ensures the existence of the $\tau$-function, defined by a
system of two first order difference equation
\begin{equation}\label{tau-def}
u_1({\bf x}) = \frac{\tau_{{\bf x}+{\bf e}_2}}{\tau_{{\bf x}}}\;,\quad
u_2({\bf x}) = \frac{\tau_{\bf x}}{\tau_{{\bf x}+{\bf e}_1}}\;.
\end{equation}
With this substitution the system (\ref{uvmap-class-again}) can be
rewritten as 
\begin{equation}\label{v-tau}
\begin{array}{l}
\ds \frac{v_1({\bf x}+{\bf e}_1)}{v_1({\bf x})} = \frac{z_1\tau_{\bf x} - \tau_{{\bf x}+{\bf e}_2}}{z_1\tau_{{\bf x}+{\bf e}_1}-\tau_{{\bf x}+{\bf e}_1+{\bf e}_2}}\;,\\
[5mm]
\ds \frac{v_2({\bf x}+{\bf e}_2)}{v_2({\bf x})} = \frac{z_2\tau_{{\bf x}+{\bf e}_2} - \tau_{{\bf x}+{\bf e}_1+{\bf e}_2}}{z_2\tau_{{\bf x}}-\tau_{{\bf x}+{\bf e}_1}}\;,\\
[5mm]
\ds \frac{v_1({\bf x})}{v_2({\bf x})} = z_2 \frac{\tau_{{\bf x}+{\bf e}_1}\tau_{{\bf x} +{\bf e}_2} - \tau_{\bf x}\tau_{{\bf x}+{\bf e}_1+{\bf e}_2}}{(z_1\tau_{\bf x} - \tau_{{\bf x} + {\bf e}_2})(z_2\tau_{\bf x} - \tau_{{\bf x}+{\bf e}_1})}\;.
\end{array}
\end{equation}
Here we have only three equations, since (\ref{uuuu}) is satisfied 
automatically by virtue of \eqref{tau-def}. 
Integrating the first two equations, one obtains
\begin{equation}
v_1({\bf x}) = \frac{\beta_{x_2}}{z_1\tau_{\bf x} - \tau_{{\bf x}+{\bf
      e}_2}}\;,\quad v_2({\bf x}) = \frac{z_2\tau_{\bf x}-\tau_{{\bf
      x}+{\bf e}_1}}{z_2\alpha_{x_1}}\;, \label{v12-tau}
\end{equation}
where ${\bf x}=(x_1, x_2)$ are the light cone coordinates,
\begin{equation}
{\bf x} = x_1{\bf e}_1+x_2{\bf e}_2\;,\quad x_1=n\;,\quad x_2=t-n\;,
\end{equation}
and $\alpha_{x_1}$, $\beta_{x_2}$ are arbitrary constants of
integration.  The third equation in (\ref{v-tau}) reduces to a single
equation for the $\tau$-function,
\begin{equation}
\tau_{{\bf x}+{\bf e}_1}\tau_{{\bf x}+{\bf e}_2} - \tau_{\bf x}\tau_{{\bf x}+{\bf e}_1+{\bf e}_2} = \alpha_{x_1}\beta_{x_2}\;.\label{liouville}
\end{equation}
This is the inhomogeneous discrete Liouville equation, involving two
arbitrary functions $\alpha_{x_1}$ and $\beta_{x_2}$. The
corresponding geometric arrangement of the $\tau$-functions 
is shown in Fig.\ref{quad2}. This is a generalization of the homogeneous
discrete Liouville equation (with $\alpha_{x_1}\equiv1$,
$\beta_{x_2}\equiv1$), previously studied in
\cite{FT1986,Hirota1987,FV1999,FKV:2001}.  It appears that the
inhomogeneity does not bring notable complications. Indeed, a
general solution to Eq.\eqref{liouville} is given by the formula 
\begin{equation}
\tau_{\bf x} = \frac{1+f_{x_1}g_{x_2}}{\phi_{x_1}\gamma_{x_2}}\;,
\label{tau-sol}
\end{equation}
where $\phi_{x_1}$, $\gamma_{x_2}$ are arbitrary functions, while 
$f_{x_1}$ and $g_{x_2}$ are determined by the following first order
difference equations  
\begin{equation}
f_{x_1+1}-f_{x_1} = \alpha_{x_1}\phi_{x_1}\phi_{x_1+1}\;,\quad
g_{x_2+1}-g_{x_2} = \beta_{x_2}\gamma_{x_2}\gamma_{x_2+1}\;.\label{fg-def}
\end{equation}
To summarize the formulae \eqref{tau-def}, \eqref{v12-tau},
\eqref{tau-sol} and \eqref{fg-def}, together with the definitions 
\eqref{v-sigma} and \eqref{uv-change}, provide the most general solution
to both the Lagrangian equations \eqref{stateq}, 
\eqref{odd-even} as well as the Hamiltonian equations of motion
\eqref{uvmap-class-again}. The solution contains $4N$ arbitrary
constants $\alpha_{x_1}$, $\beta_{x_2}$, $\phi_{x_1}$,
$\gamma_{x_2}$ ($x_1,x_2=1,\ldots,N$),\ whose number precisely coincides with the dimension of
the phase space of our classical discrete evolution system \eqref{cevol}.

\section{Conclusion}
In this paper we have shown that the theory of integrable maps (and more
specifically the so-called Yang-Baxter maps) can
be naturally included as a part of the modern theory of 
integrable system, based on the theory of quantum groups, the Quantum
Inverse Problem Method and their classical counterparts. The
considerations apply to both quantum and classical theories (the latter
arise as the quasiclassical limit of the quantum case). One of the
main advantages of our algebraic approach is that the resulting 
discrete integrable evolution systems automatically possess
meaningful Hamiltonian structures. In addition, the problem of
quantization of the Yang-Baxter maps appears to be solved from the
very beginning. 

We have illustrated the entire scheme on the example on the quantized
Lie algebra $U_q(sl(2))$ (and its Poisson algebra limit for the classical
case). Completely parallel presentations of the quantum and classical
cases are given in Sect.~2 and Sect.~3 respectively.  In particular, a
general solution the classical discrete Liouville equations arising in this
context (both in the
Lagrangian and Hamiltonian forms) is given in Sect.~3.11.  

Although the above considerations were mainly restricted to the case of
$U_q(sl(2))$, our approach is rather general and can be applied to any
quantized Lie (super) algebra as well as their affine extensions. We hope to
consider some of these problems in future publications.

\section*{Acknowledgements}
We thank Ludwig Faddeev, Rinat Kashaev and Joerg Teschner 
for important comments and Nicolai
Reshetikhin for alerting us about Sklyanin's paper
\cite{Sklyanin:1988}. 
We also
thank Sergei Khoroshkin and Zengo Tsuboi for numerous 
fruitful discussions and collaboration
on related projects.

%\newpage
\app{Additional properties of the $\Rb$-matrix}
\noindent
Here we present some additional relations connected with the $\Rb$-matrix
\eqref{Rfact}. The non-compact quantum dilogarithm \eqref{noncom}
(sometimes called the Barnes double-sine function) is a 
unique solution of the functional equation
\be\label{dilog-difference}
\log\varphi(z-\ii\frac{\bb}{2}) - \log\varphi(z+\ii\frac{\bb}{2})=\log(1+\EXP^{2\pi\bb z}) 
\ee
provided that $\varphi(z)$ is analytic in the strip
\be
-\textrm{Re}(\bb)\leq \textrm{Im}(z)\leq \textrm{Re}(\bb)\;.
\ee
It satisfies the relations
\be\label{propphi}
\varphi(z)\varphi(-z)=\EXP^{\ii\pi z^2-\ii\pi(1-2\eta^2)/6}\;,\quad \left.\varphi(z)\right|_{z\to -\infty} \to 1\;,\quad
\varphi(z)^*=\frac{1}{\varphi(z^*)}\;.
\ee
Define also another function 
\be
\Phi(z)=\exp\left(\frac{1}{8}\int_{\mathbb{R}+\ii 0}\frac{\EXP^{-2\ii z w}}{\sinh(w\bb)\sinh(w/\bb)\cosh(2\eta w)}\frac{dw}{w}\right)\;.
\ee
which possesses the properties
\be\label{PropPhi}
\Phi(z)\Phi(-z)=\EXP^{\ii\pi z^2/2-\ii\pi(1-8\eta^2)/12}\;,\quad \left.\Phi(z)\right|_{z\to-\infty}\to 1\;.
\ee

The Boltzmann weights for the Faddeev-Volkov model (which is related
to the quantized affine algebra $U_q(\widehat{sl}_2)$) are defined
as \cite{BMS07a}, 
\begin{equation}
W_\alpha(s)=\frac{1}{F_\alpha}\EXP^{2\pi\alpha s} \frac{\varphi(s+\ii\alpha)}{\varphi(s-\ii\alpha)}\;,\quad \overline{W}_\alpha(s)=W_{\eta-\alpha}(s)\;,
\end{equation}
where
\begin{equation}
F_\alpha=\EXP^{\ii\pi\alpha^2+\ii\pi(1-8\eta^2)/24}\Phi(2\ii\alpha)\;.
\end{equation}
It is useful to note that $W_\alpha(s)=W_\alpha(-s)$. 
Consider also a simple limit,
\begin{equation}
s\to s-K\;,\quad \ii\alpha\to\ii\alpha -K\;,\quad K\to \infty
\end{equation}
In this case, from \eqref{propphi} and \eqref{PropPhi} it follows that
\begin{equation}
W_\alpha(s)\;\to\; \EXP^{-\ii\pi K^2 + 2\pi\ii Ks} V_\alpha(s)\;,\qquad
\overline{W}_\alpha(s)\;\to\; \EXP^{\ii\pi K^2 - 2\pi\ii Ks} \overline{V}_\alpha(s)\;,
\end{equation}
where $V_\alpha(s)$ and $\overline{V}_\alpha(s)$ are defined in \eqref{Vdef}.

The weights $W_\alpha(s)$ and $\overline{W}_\alpha(s)$ satisfy the
following star-triangle relation \cite{FV95,BMS07a}, 
\begin{equation}\label{fvstr}
\int_{\mathbb{R}}\,d\sigma \,\overline{W}_{\alpha}(a-\sigma)W_{\alpha+\beta}(c-\sigma)\overline{W}_\beta(b-\sigma) = 
 W_\beta(a-c) \overline{W}_{\alpha+\beta}(a-b) W_{\alpha}(c-b)
\end{equation}
which admits several interesting limits.
For example, if
\begin{equation}
a\to a-K\;,\quad c\to c-K\;,\quad \ii\alpha\to\ii\alpha-K\;,\quad K\to\infty\;,
\end{equation}
then the singular terms in \eqref {fvstr} cancel out 
and the relation reduces to 
\begin{equation}\label{str1}
\int_{\mathbb{R}}\,d\sigma\, \overline{V}_{\alpha}(a-\sigma)V_{\alpha+\beta}(c-\sigma)\overline{W}_\beta(b-\sigma) = 
 W_\beta(a-c) \overline{V}_{\alpha+\beta}(a-b) V_{\alpha}(c-b)
\end{equation}
A similar limit,
\begin{equation}
a\to a+K\;,\quad c\to c+K\;,\quad \ii\alpha\to\ii\alpha-K\;,\quad K\to\infty\;,
\end{equation}
yields
\begin{equation}\label{str2}
\int_{\mathbb{R}}\, d\sigma\, \overline{V}_{\alpha}(\sigma-a)V_{\alpha+\beta}(\sigma-c)\overline{W}_\beta(b-\sigma) = 
 W_\beta(a-c) \overline{V}_{\alpha+\beta}(b-a) V_{\alpha}(b-c)\;.
\end{equation}
These relations can be recursively used to prove the Yang-Baxter
equation \eqref{ybe-integ} given in the main text.

It is interesting to study the classical limit of the above relations
\eqref{str1} and \eqref{str2}. 
The classical limits of functions $V$ and $\overline{V}$ 
is given by (\ref{Vcl}). Similar limits of $W$ and $\overline{W}$ are given by
\begin{equation}
W_{\frac{\ii a}{2\pi\bb}}\Big(\frac{{\sigma}}{2\pi\bb}\Big) =
\exp\left\{\frac{\ii}{2\pi\bb^2}\Lambda_{a}({\sigma})+O(1)\right\}\,,\qquad
\overline{W}_{\frac{\ii{a}}{2\pi\bb}}\Big(\frac{{\sigma}}{2\pi\bb}\Big)=
\exp\left\{\frac{\ii}{2\pi\bb^2}\overline{\Lambda}_{a}({\sigma})+O(1)\right\}\;,
\end{equation}
where
\begin{equation}
\frac{\partial\Lambda}{\partial\sigma}=\log\frac{e^\sigma+e^\alpha}{1+e^{\sigma+\alpha}}\;,\qquad\qquad 
\frac{\partial\overline\Lambda}{\partial\sigma}=\log\frac{e^\sigma-e^{-\alpha}}{1-e^{\sigma-\alpha}}\;.
\end{equation}
Then, for instance, the stationary point of the integral (\ref{str2}) 
in the limit $\bb\to 0$ is determined by the relation 
\begin{equation}
\frac{\partial}{\partial\sigma}\left(\overline{\lambda}_{\alpha}(\sigma-a) + \lambda_{\alpha+\beta}(\sigma-c) + \overline{\Lambda}_{\beta}(\sigma-b)\right)=0\;,
\end{equation}
which is equivalent to 
\begin{equation}
\log(e^{\sigma-a}-e^{-\alpha}) - \log(e^{\sigma-c}+e^{-\alpha-\beta}) +\log\frac{e^{\sigma-b}-e^{-\beta}}{1-e^{\sigma-b-\beta}} = 0\;.
\end{equation}
The last relation has the structure of the Adler-Bobenko-Suris \cite{AdlerBobenkoSuris} tree-legs equation 
\begin{equation}
\psi(x_0,y_0;\alpha_0)-\psi(x_0,y_1;\alpha_1)=\varphi(x_0,x_1;\alpha_0,\alpha_1)
\end{equation}
corresponding to the $H_3$ system in their classification.

\app{Relation to $R$-matrix for quantum Teichm\"uller theory}
The $\Rb$-matrix (\ref{Ropfact}) coincides with the inverse of $\Rb$-matrix (44) from \cite{Kashaev2000}. Details are the following.
Moving quadratic exponents in (\ref{Ropfact}) to the right, one could obtain the left hand side of the following identity:
\begin{equation}
\ii \EXP^{-\ii\pi (\xop_1-\xop_2)^2} \EXP^{-\ii\pi (\pop_1^2+\pop_2^2)} \EXP^{-\ii\pi(\xop_1-\xop_2)^2} \mathbf{P}_{12} = \EXP^{-2\pi\ii\pop_1\pop_2}\;.
\end{equation}
This identity can be easily proven by consideration of a kernel or of a normal symbol of the left and right hand sides and performing the Gauss integration.

Using the Pentagon identity (\ref{pentagon}) twice in (\ref{Ropfact}), one can then bring 
$\Rb_{12}$ to the form 
\begin{equation}
\Rb_{12} = \mathbf{F}_{12}^{} \varphi(\ii(\beta_1-\alpha_2)+\xop_1-\xop_2) \mathbf{F}_{12}^{-1} \EXP^{-2\pi\ii\pop_1\pop_2}\;,
\end{equation}
where
\begin{equation}
\mathbf{F}_{12}^{} = \varphi(\ii(\alpha_1-\beta_1) + \pop_1) \varphi(\ii(\alpha_2-\beta_2)-\pop_2)
\;.
\end{equation}
This form of $\Rb_{12}$ literally coincides with $\Rb^{-1}$, formula (44) from \cite{Kashaev2000}, after indentifying corresponding variables.
The authors are grateful to R. M. Kashaev for this observation.

%\addcontentsline{toc}{chapter}{References}
%\bibliography{total33,maps}
%\bibliographystyle{vvb-bibstyle}
%\bibliographystyle{utphys1}

\newcommand\oneletter[1]{#1}\def\cprime{$'$} \def\cprime{$'$}
\providecommand{\href}[2]{#2}\begingroup\raggedright\endgroup 

\end{document}